\documentclass[10pt, superscriptaddress, twocolumn,showpacs,showkeys,amsmath,amssymb,nofootinbib]{revtex4-1}

\usepackage{amsmath, amssymb, latexsym, verbatim}
\usepackage{graphicx}
\usepackage{dcolumn}
\usepackage{bm}
\usepackage[dvips]{color}

\newcommand{\di}{\partial}

\newcommand{\f}{\frac}

\newcommand{\nin}{\noindent}

\newcommand{\D}{\textup{d}}
\newcommand{\El}{\mathcal{L}}

\newcommand{\prm}{\prime}

\newcommand{\sabs}[1]{\ensuremath{\lvert #1 \rvert}}

\begin{document}

\title{Unifying relativity and classical dynamics}

\author{Mozafar Karamian} \email{karamian@ymail.com} \noaffiliation

\author{Mahdi Atiq} \email{mma\_atiq@yahoo.com}
\affiliation{Foundations of Physics Group, School of Physics, Institute for Research in Fundamental Sciences (IPM), Tehran 19395-5531, Iran}

\author{Fatemeh Najdat} \email{fa\_najdat@yahoo.com}\noaffiliation

\author{Mehdi Golshani} \email{mehdigolshani@yahoo.com} \affiliation{Foundations of Physics Group, School of Physics, Institute for Research in Fundamental Sciences (IPM), Tehran 19395-5531, Iran}
\affiliation{Department of Physics, Sharif University of Technology, Tehran 11365-8639, Iran}

\begin{abstract} 
\nin Relativity and classical dynamics, as defined so far, form distinct parts of classical physics and are formulated based on independent principles. We propose that the formalism of classical dynamics can be considered as the theoretical foundation of the current theory of relativity and may be employed for exploring possibilities beyond the current theory. We show that  special-relativistic kinematics, including universality of the speed of massless particles relative to inertial frames, is a consequence of the formalism of classical dynamics, with no assumptions other than spacetime point transformations and Euclidean geometry of space in inertial frames. We discuss that energy-independent velocity is a general concept in classical dynamics, applicable even to massive objects, in appropriate canonical coordinates. The derivation of Lorentz symmetry is inherently local and allows the speed of massless particles (relative to local inertial frames) to vary with space and time globally, which may provide a theoretical foundation for variable speed of light cosmology. We obtain no kinematical scales other than the light-speed, specially no scale of energy or momentum as has been suggested in some quantum gravity investigations. We argue that this is a consequence of spacetime \emph{point} transformations making the momentum space linear, and a possible second scale must require \emph{non-point} transformations as a necessary condition, which seems compatible with the notion of relative locality in curved momentum space.
\end{abstract}



\maketitle

\section{Introduction}
\nin Classical physics, as has been developed and formulated so far, consists of two basic parts, relativity and classical dynamics, which are defined as distinct theoretical frameworks and are formulated using independent principles. 

Relativity, as the more fundamental theory, is defined as the theory of properties and dynamics \emph{of} spacetime, distinct from theories that describe properties and dynamics of matter \emph{in} spacetime, and spacetime is defined in the theory as an arena on which in principle any theory of matter, classical or quantum, can be defined. Apart from the Einstein field equations describing the dynamics of spacetime, the theory is based on the principles of equivalence of inertial frames, universality of the speed of massless particles relative to inertial frames, inertia-gravity equivalence, and general covariance.

Classical dynamics, on the other hand, is considered as a restricted (non-quantum) framework of dynamics, which by itself can even be non-relativistic, and is formulated based on the action principle. It, therefore, is considered as one of the possible dynamical frameworks that can be developed on the relativistic spacetime.

In other words, relativity defines the fundamental kinematics upon which theories of dynamics are developed, although the kinematics itself has a dynamical character (in the presence of gravity).  

Therefore, according to the current definition of the theory, while spacetime interacts with matter, no theoretical relationship between properties of spacetime and the classical or quantum nature of the dynamics of matter is assumed. Accordingly, although the theory deals with motions of particles and reference frames in the classical sense and using spacetime geodesics whose equations are mathematically equivalent to some Lagrange equations, the pseudo-Riemannian structure of the underlying manifold is not theoretically related to Lagrangian formalism. It is assumed as a characteristic or defining property of gravitational field. 

This theoretical independence between relativity and classical dynamics, however helpful in developing different dynamical frameworks, including quantum ones, does not seem theoretically satisfactory in some aspects. Relativity in its current form is a classical (non-quantum) theory in every aspects. Inertial frames, which have a central role in the theory, are indeed rest frames of objects which move according to the laws of classical dynamics, with definite trajectory, and time, length, velocity, mass, energy, momentum and so on in those frames are classical concepts developed in terms of the classical behavior of objects. The theory fundamentally stands upon concepts taken from classical dynamics, and the relativistic kinematics which is used as a foundation even in quantum field theory gets its meaning originally from the classical behavior of objects.

But, apart from the classical \emph{nature} of the theory, the mathematical formalism of the theory has essential features in common with the formalism of classical dynamics, which should be considered as indicating the requirement of a theoretical unification between them. As mentioned above, geodesic equations can be expressed in the form of Lagrange equations, and the action principle becomes equivalent to the condition of shortest path (minimal proper time) in this case. Even without presuming anything about Einsteinian relativity with its specific principles mentioned above, Lagrange equations are intrinsically invariant under arbitrary changes of coordinate system in the configuration space, which is practically the requirement of general covariance in relativity. More importantly, any Lagrangian which is homogeneous of degree one in its velocity arguments provides a measure of length and a metric tensor on the configuration space \cite{c: Rund59}. This means that Lagrangian formalism, by its intrinsic properties, can be employed for defining a Finsler or Riemann metric on the configuration space, and therefore the concept of metric can be introduced very naturally into the theory starting from Lagrangian formalism. Such intrinsic properties and capabilities of the Lagrangian formalism can be considered as theoretical evidences for a unified point of view regarding the current theory of relativity and classical dynamics. In other words, the mathematical formalism of classical dynamics is so powerful that one may consider this formalism not just as a framework of dynamics (a \emph{restricted} one), but even as the foundation of the theory of relativity, and specifically and importantly, as the theoretical foundation of relativistic kinematics.

In the following, we treat classical dynamics not just as a theoretical framework governing classical motions of particles and reference frames, but also as the theoretical origin of relativistic kinematics. Our basic intention is to demonstrate that one can start completely and fundamentally from the intrinsic properties of Lagrangian formalism (without making any assumption about the functionality of the Lagrangian) instead of the above mentioned principles of Einsteinian relativity and still end up with relativistic kinematics, and those principles should in fact be considered as consequences of the Lagrangian formalism --- except for the principle of inertia-gravity equivalence which has to be taken as a purely empirical fact of nature, without which even definition and existence of inertial frame would be impossible. 

Underlying all our derivations and conclusions is a theorem (proved in section \ref{S: Invariance of action}) which provides the basis for developing the theory of relativity using Lagrangian formalism and states that an \emph{ordinary} action function defined by $S(t, q^i) = \int_{(t_0, q_0)}^{(t, q)} L \D t$ is automatically a scalar function with respect to point transformations of the \emph{extended} configuration space, with coordinates $t$ and $q^i$. (Here, $L$ is an \emph{ordinary} Lagrangian, scalar only with respect to point transformations of the configuration space with coordinates $q^i$, and the integral is taken along an extremal trajectory.)

The most fundamental ingredient of Einsteinian relativity, i.e. the relativity of time, is completely independently explained as incompatibility of absolute time with the intrinsic invariance of action and even with the Euclidean geometry of space in inertial frames. The impossibility of a well-defined inertial rest frame for massless particles is also concluded, very straightforwardly, from the invariance of action. 

We present a detailed derivation of Lorentz symmetry and universality of the speed of massless particles as consequences of the invariance of action and Euclidean geometry of space. 

The equivalence between inertial frames is not required to be assumed as an independent or fundamental principle besides the Lagrangian formalism. Instead, it can be considered as a systematic consequence of the formalism. The connection between inertiality of the frame and the Euclidean geometry of space (more accurately speaking, equivalence between them) is also explained using Lagrangian formalism.\footnote{We should emphasize that we do not assume any form for Lagrangian in our approach. The mathematical form of the Lagrangian of free particles relative to inertial frames, for example, is obtained from the invariance of action and the Euclidean geometry of space.}

As a supporting theoretical evidence for the fundamental role of classical dynamics in the foundations of relativity, we show that \emph{energy-independent velocity}, which is attributed only to massless particles in the Einsteinian formulation of relativity, is actually a more general concept in classical dynamics for the generalized meaning of the term \emph{velocity}, and is applicable to massive and massless systems, in appropriate canonical coordinates. An example of such a velocity in a massive system is the energy-independent angular velocity of harmonic oscillator. We show that this similarity with the energy-independent speed of light is not accidental, but is a consequence of the vanishing of Lagrangian, and appears in any system in which the Lagrangian vanishes during the evolution of the system. The concept of energy-independent velocity is not canonically-invariant in general, as is observed from the example of harmonic oscillator which has an energy-\emph{dependent} linear velocity, but an energy-\emph{independent} angular velocity. The linear energy-independent velocity of massless particles in the ordinary space is, however, observer-independent, as far as the transformations between observers are assumed to be point transformations of spacetime. The energy-independent velocity, by itself, completely characterizes the motions of massless particles in arbitrary spacetime coordinates, without imposing the assumption of an absolutely constant speed relative to local inertial frames throughout the particle trajectory.

The presented approach to the foundations of relativity and relativistic kinematics has consequences relevant to some investigations in cosmology and quantum gravity. The discussed derivation of Lorentz symmetry from invariance of action is inherently and fundamentally local in two senses. It is fundamentally local in the sense that it is based on an assumption of locality, the spacetime point transformations. We argue that this makes the momentum space a linear space and therefore eliminates the possibility of a kinematical scale of energy or momentum suggested in the so-called doubly special relativity approach in quantum gravity \cite{c: DSR01,c: DSR02,c: DSR03,c: DSR04}. Such a scale, if exists at all, would require \emph{non-point} transformations of spacetime, which seems consistent with the notion of relative locality in a curved momentum space \cite{c: DSR08,c: DSR09}. The derivation is also inherently local in the sense that the symmetry is derived for local inertial frames and therefore allows the speed of massless particles (relative to local inertial frames) to vary with space and time globally. This can provide a theoretical foundation for variable speed of light cosmology which has previously been considered by other researchers \cite{c: VSL01,c: VSL02,c: VSL03}. What supports the idea of variable speed of light is the fact that motions of massless particles in spacetime are most fundamentally characterized by energy-independent velocity, not by the stronger condition of an absolutely constant speed (relative to local inertial frames) in the entire spacetime. 

Derivation of relativistic kinematics as a consequence of classical dynamics has implications for the status of Lorentz symmetry in general relativity and the relationship between kinematics and dynamics. According to general relativity, space and time are fundamentally dynamical concepts, but nevertheless, the most fundamental building block of the theory, Lorentz symmetry, is assumed in the theory as a purely-kinematical phenomenon and an intrinsic property of space and time in inertial frames, with no dynamical character or origin whatsoever. Our approach, while retains (locally) the kinematical role of Lorentz symmetry, but gives it a dynamical origin by deriving it as a consequence of classical dynamics. This dynamical origin is also meaningful in the sense that the symmetry is a symmetry of space and time, and classical dynamics concerns evolution of systems in space with time.

Unification of relativity with classical dynamics and providing a foundation for Lorentz symmetry in the latter is closer than the purely-kinematical approach of special relativity to the essential dynamical nature of space and time in general relativity and adds another aspect to that dynamical nature in the formalism of the theory: even properties of spacetime in inertial frames, yielding the concept of Minkowski spacetime, has a dynamical origin, and that origin is classical dynamics.

This point of view does not revoke the kinematical applications of Lorentz symmetry, but implies a reciprocal relationship between kinematics and dynamics. Not only theories of dynamics are built upon kinematical assumptions, those assumptions themselves can be consequences of some dynamical frameworks.

We present this work, not just as a unification of the two basic parts of 
classical physics, but also as a new approach to the foundations of relativity and in fact a re-definition of this theory 
and we hope that it can provide a theoretical framework upon which one can explore 
theoretical possibilities beyond the current theory of relativity.

\section{A motivation for incorporating Lorentz symmetry into the formalism of classical mechanics}\label{S: Motivation}
\nin According to the theory of special relativity, Lorentz symmetry is a purely-kinematical symmetry and a characteristic property of space and time in inertial frames, independent of anything dynamical, and therefore is not assumed as a consequence of any dynamical framework, including classical dynamics. This point of view, which has originally been taken from consideration of Maxwell's fields, is specially suitable for field theories (classical or quantum), where Lorentz symmetry is simply implemented as a kinematical symmetry of space and time, independent of the dynamics of the fields. But, the same point of view has been applied also to classical particle dynamics, and is clearly observed in the derivations of relativistic mechanics. However, the latter is questionable, since, as is discussed below, Lorentz symmetry might meaningfully be considered as a consequence of the classical particle dynamics applied to the motions of the inertial frames themselves (more accurately, to the motions of particles sitting in inertial frames).

In one derivation of the proper Lagrangian for relativistic particles \cite{c: LL}, which is among the simplest ones, in order to obtain the Lagrangian, the action $S$ is identified (up to a constant coefficient) with the proper time $\tau$ of the particle, such that one has $S = \int L \D t = - mc^2 \int \D \tau = - mc^2 \int \gamma^{-1} \D t$, which immediately yields $L = - mc^2 \gamma^{-1}$. Here, as usual, $\gamma = (1 - v^2/c^2)^{-1/2}$ and $m$ is particle's mass.

Here, the time dilation relation $\D \tau = \gamma^{-1} \D t$ is provided from somewhere \emph{unrelated to Lagrangian formalism}, that is, from Lorentz transformation which according to the current theory is based on the universality of the speed of light as a fundamental kinematical property of space and time with no foundation in any framework of dynamics including classical dynamics.

Let us now consider the famous equation $\D S = p_i \D q^i - H \D t$, which is a direct consequence of the definition of action ($S = \int L \D t$) regardless of the assumed transformation between inertial frames, Galileo or Lorentz (here $t$ is the time parameter of observer and $L$ is an \emph{ordinary} Lagrangian which is scalar only with respect to spatial coordinates.). This equation, therefore, requires no modification in special relativity. In this case, since action is defined as a Lorentz-invariant quantity, one has $\D S = - mc^2 \D \tau$ in the rest frame of the particle. However, if one derives from the above mentioned Lagrangian the expressions of momentum and energy, $p_i = m \gamma \delta_{ij} v^j$ and $H = mc^2 \gamma$, and substitutes them in the above equation for $\D S$, one obtains $\D \tau = \gamma (\D t - \delta_{ij} v^j \D q^i)$, which is one of the equations of Lorentz transformation. 

We observe that if we start from Lorentz transformation, obtain the time dilation relation and from which the Lagrangian, and derive the expressions of energy and momentum, then the equation $\D S = p_i \D q^i - H \D t$ takes us back to the Lorentz transformation which was employed for obtaining the Lagrangian at the first place. As a consequence, one may say that this equation is at least equivalent to one equation of Lorentz transformation in the special-relativistic case. 

Now, suppose that we did not know the Lorentz transformation, but we somehow knew that $S$ is an invariant and we could obtain the free particle Lagrangian $L = -E_0 \gamma^{-1}$ purely from the Lagrangian formalism without employing Lorentz transformation. Then, we could obtain the above mentioned equation of Lorentz transformation from the above-mentioned equation of the action. In other words, instead of assuming Lorentz transformation and obtaining a proper Lagrangian for classical particles, we may conversely obtain the Lorentz transformation from the Lagrangian formalism by applying it to particles sitting in inertial frames and representing motions of such frames, without starting from the assumption of universal light-speed as a fundamental kinematical principle. In fact, in this case, universality of the speed of massless particles relative to inertial frames must appear as a consequence of the invariance of action in the Lagrangian formalism, and taking the universal light-speed or Lorentz symmetry as fundamental principle would be redundant in Lagrangian formalism after taking $S$ as an scalar.

The above mentioned observation, alongside other theoretical elements existing in relativity and classical dynamics mentioned in the previous section,  motivates us to suggest that relativistic kinematics could be considered as a consequence of classical dynamics, and that Lorentz symmetry could be incorporated into the formalism of classical dynamics in a fundamental way. 

This attitude regarding Lorentz symmetry when dealing with classical dynamics is meaningful since Lorentz symmetry is a symmetry among inertial frames, and inertial frames are just rest frames of objects which move under the laws of classical (non-quantum) dynamics, with definite trajectory. The concepts of space and time, which are the subjects of Lorentz symmetry, are classical concepts taken from the behavior of objects in the macroscopic world, governed by the laws of classical dynamics. Therefore, unlike the situation of field theories, when it comes to particle dynamics, which includes motions of reference frames as well, fundamental separation of spacetime symmetry from the dynamics of particles does not seem theoretically justified. In other words, Lorentz symmetry, which has a kinematical role in field theories, can itself be a consequence of classical particle dynamics.

In the following section, we discuss some fundamentals based on the intrinsic properties of Lagrangian formalism which provide the foundation for developing the current theory of relativity within that formalism.

\section{Fundamentals from Lagrangian formalism}\label{S: Fundamentals}

\subsection{Invariance of action}\label{S: Invariance of action}
\nin In this subsection, we prove the following theorem which underlies all our conclusions and derivations in the entire paper:

\emph{An action function defined by $S(t, q^i) = \int_{(t_0, q_0)}^{(t, q)} L \D t$, where the Lagrangian $L$ is scalar only with respect to point transformations of configuration space with coordinates $q^i$ and the integral is taken along an extremal trajectory from fixed $t_0$ and $q_0$, is automatically a scalar function with respect to point transformations of the \textup{extended} configuration space with coordinates $t$ and $q^i$.}\\

Using an ordinary Lagrangian $L(t, q^i, v^i)$, where $v^i = \f{\D q^i}{\D t}$ and $t$ is the time parameter of observer, one can define an action function 

\begin{equation}\label{S: S = int L dt}
	S(t, q; t_0, q_0) = \int_{(t_0, q_0)}^{(t, q)} L \D t,
\end{equation}

\nin where the integral is taken along an extremal trajectory connecting the bounds of the integral in the \emph{extended} configuration space of $t$ and $q^i$. If one keeps $(t_0, q_0^i)$ fixed, this can be considered as a function $S(t, q^i)$ in that space. Since $L$ is a scalar with respect to the coordinate system $\{q^i\}$ of the \emph{configuration} space, $S$ too is a scalar in that respect. We, however, show that the same function $S$ is a scalar even with respect to arbitrary coordinate systems of the extended configuration space. This is done by showing that, because of the intrinsic properties of Lagrange equations, ordinary Lagrange equations in the configuration space are always equivalent to some other Lagrange equations which hold in the extended configuration space. Showing the validity of the theorem requires the following steps.

We begin by recalling that the equation $\f{\D (-H)}{\D t} = \f{\di L}{\di t}$, where $H$ is the Hamiltonian corresponding to $L$, is a consequence of Lagrange equations $\f{\D p_i}{\D t} = \f{\di L}{\di q^i}$. If one defines $q^0 = t$ and $p_0 = - H$, these equations can be written altogether in the form

\begin{equation}\label{eq: Lagrange eqs of L}
	\f{\D p_\mu}{\D t} = \f{\di L}{\di q^\mu},
\end{equation}

\nin remembering that one of these equations (the one corresponding to $\mu = 0$) is not independent of the others. Intrinsic properties of Lagrange equations allow us to recast these equations in another form in which the observer time parameter $t$ is treated as a coordinate like the spatial coordinates $q^i$, as follows.

It is well-known \cite{c: Gold} that if one defines a new Lagrangian 

\begin{equation}\label{eq: El}
	\El (q^\mu, \dot{q}^\mu) = L(t, q^i, \f{\D q^i}{\D \tau} (\f{\D t}{\D \tau})^{-1}) \f{\D t}{\D \tau},
\end{equation}

\nin where $\tau$ is a new and arbitrary parametrization of the trajectory in the extended configuration space and $\dot{q}^\mu = \f{\D q^\mu}{\D \tau}$, one obtains from the ordinary Lagrange equations, i.e. equations \eqref{eq: Lagrange eqs of L}, that

\begin{equation}\label{eq: Lagrange eqs of El}
	\f{\D p_\mu}{\D \tau} = \f{\di \El}{\di q^\mu},
\end{equation}

\nin in which, $p_\mu$ expresses the energy and momentums defined in terms of $L$, i.e. $p_i = \f{\di L}{\di v^i}$ and $p_0 = - H = L - p_i v^i$, where $v^i = \f{\D q^i}{\D t}$. Incidentally, one obtains the equalities $p_\mu = \f{\di \El}{\di \dot{q}^\mu}$ by taking the derivatives of $\El$ with respect to $\dot{q}^\mu$, which make equations \eqref{eq: Lagrange eqs of El} genuine Lagrange equations in the extended configuration space.

The new Lagrangian $\El$ is homogeneous of degree one in $\dot{q}^\mu$ and, as a consequence, one of the equations \eqref{eq: Lagrange eqs of El} is dependent on the others \cite{c: Gold,c: Rund66}, as expected from equations \eqref{eq: Lagrange eqs of L}.

Similarly and conversely, if one assumes a Lagrangian $\El(q^\mu, \dot{q}^\mu)$ which is homogeneous of degree one in the velocity variables $\f{\D q^\mu}{\D \tau}$ in the extended configuration space, one can define a Lagrangian $L(t, q^i, \f{\D q^i}{\D t})$ such that

\begin{equation}\label{eq: L}
	L(t, q^i, \f{\D q^i}{\D t}) \equiv \El (q^\mu, \f{\D q^\mu}{\D \tau}) (\f{\D t}{\D \tau})^{-1} = \El(t, q^i, 1, \f{\D q^i}{\D t}). 
\end{equation}

\nin One can then recover from equations \eqref{eq: Lagrange eqs of El} the equations \eqref{eq: Lagrange eqs of L} in which $p_\mu$, now defined by $\f{\di \El}{\di \dot{q}^\mu}$, reduce to $p_i = \f{\di L}{\di v^i}$ and $p_0 = L - p_i v^i$. Here, one singles out one of the coordinates, denoted by $t$, to be used as a parametrization of trajectory in the extended configuration space. This process is \emph{coordinate system-dependent} and, as a consequence, although $\El$ is a scalar in the extended configuration space, the resulting $L$ is specific to the chosen parameter $t$ and is scalar only with respect to the coordinates $q^i$.

So, one can obtain equations \eqref{eq: Lagrange eqs of El}, which hold in the extended configuration space, from equations \eqref{eq: Lagrange eqs of L}, which hold in the configuration space, and vice versa. Mathematically, this means that those two sets of equations are equivalent.

This mathematical equivalence is independent of our physical assumption about the (absolute or relative) nature of time. It is only because of the intrinsic properties of Lagrange equations that the ordinary Lagrange equations \eqref{eq: Lagrange eqs of L} are equivalent to another set of Lagrange equations which are defined in the extended configuration space, allowing the time parameter of the observer, used in equations \eqref{eq: Lagrange eqs of L}, to be chosen arbitrarily using the equivalent equations \eqref{eq: Lagrange eqs of El}. This means that the assumption of absolute time is not a requirement of Lagrangian formalism and it is even an unnecessary assumption for the formalism.

A crucial consequence of the equivalence of equations \eqref{eq: Lagrange eqs of L} and \eqref{eq: Lagrange eqs of El} is that the \emph{ordinary} action function, defined by \eqref{S: S = int L dt}, automatically becomes a scalar of the extended configuration space. This simply follows from the equality $S = \int L \D t = \int \El \D \tau$ using \eqref{eq: El} or \eqref{eq: L}, and proves the above mentioned theorem.

So, mathematically speaking, regardless of our assumption about the physical nature of time, the invariance of action in the extended configuration space, where the time parameter of observer is treated as a coordinate, is a consequence of the properties of the Lagrange equations.

\subsection{Connection between rest energy and rest frame}\label{S: Rest energy - Rest frame}
\nin From the invariance of action and Hamilton-Jacobi equations ($H = - \f{\di S}{\di t}$, $p_i = \f{\di S}{\di q^i}$), it follows that energy and momentum of a particle in two different reference frames are related to each other by $p^\prm_\mu = p_\nu \f{\di q^\nu}{\di q^{\prm \mu}}$. Now, suppose that the particle is at rest in one inertial frame ($p^\prm_i = 0$) and its rest energy is zero ($-p^\prm_0 = 0$). That leads to $\f{\di q^\nu}{\di q^{\prm \mu}} p_\nu = 0$. This means that either we have $p_\nu = 0$ (for all components) or the transformation matrix $\f{\di q^\nu}{\di q^{\prm \mu}}$ is singular. In the first case, it means that $p_\nu$ is zero in all arbitrary frames, which means that the particle does not exist at all! In the second case, it means that, if the observer's rest frame is well-defined, the particle does not have a well-defined rest frame.

Therefore, invariance of action implies that a particle with a zero rest energy can not have a well-defined inertial rest frame. Equivalently, a particle which has a well-defined inertial rest frame has necessarily a nonzero rest energy.\footnote{This conclusion, of course, does not include situations where particle is at rest relative to a frame but its canonical momentum is nonzero in that frame, as is the case for charged particles in magnetic fields. It, however, includes free particles in local or global inertial frames.}

It must be noted that these conclusions are consequences of invariance of action and are obtained before deriving Lorentz transformation explicitly. They show the physical foundation of that transformation in Lagrangian formalism and the fundamental role of the invariance of action in this regard.

\subsection{Inconsistency of absolute time with invariance of action}
\nin Let us assume that both $t$ and $S$ remain invariant in going from one inertial frame to another. This means that we have $S^\prm(t, q^\prm(t, q^j)) = S(t, q^j)$. Then, from Hamilton-Jacobi equations one obtains $p_j = p^\prm_i \f{\di q^{\prm i}}{\di q^j}$. This implies that if a particle is at rest in one inertial frame ($p^\prm_i = 0$), it will be at rest in every other frame, regardless of how the frames move relative to each other. This meaningless result shows that the assumption of absolute time is inconsistent with the intrinsic invariance of action in the extended configuration space.

\subsection{Metric in Lagrangian formalism, and inconsistency of Newtonian mechanics with Euclidean geometry of space in inertial frames}\label{S: metric}
\nin According to the definition of Finsler metric \cite{c: Rund59}, a Lagrangian which is homogeneous of degree one in the velocity variables provides a measure of length and a metric tensor on the configuration space. Since the Lagrangian $\El$ defined in \eqref{eq: El} has such a property, it defines a metric tensor in the extended configuration space in the form

\begin{equation}\label{eq: Metric}
g_{\mu \nu} (q^\lambda, \dot{q}^\lambda) = \f{1}{2 \alpha_0^2}  \f{\di^2}{\di \dot{q}^\mu \di \dot{q}^\nu} \El^2 (q^\lambda, \dot{q}^\lambda),
\end{equation}

\nin where $\alpha_0$ is a \emph{nonzero} constant introduced for fixing units. (This constant is proportional to the minus rest energy in the case of massive particles.)

From this definition, it is obvious that not every Lagrangian leads to a Euclidean metric for \emph{space}. Let us examine the Newtonian Lagrangian $L = \f{1}{2} m \delta_{ij} v^i v^j$. Here, $\delta_{ij}$ has been taken as the Euclidean geometry of space. However, using the corresponding $\El$, which is $\El = \f{1}{2} m \delta_{ij} \dot{q}^i \dot{q}^j \dot{t}^{-1}$, and the consequent metric from \eqref{eq: Metric}, we find

\begin{equation}
g_{ij} = \f{m^2}{\alpha_0^2} \{ \delta_{ik} \delta_{jl} v^k v^l + \f{1}{2} (\delta_{kl} v^k v^l) \delta_{ij} \},
\end{equation}

\nin which differs from $\delta_{ij}$ and is velocity-dependent. By \emph{ad hoc} adding a rest energy $E_0$ to the Lagrangian, an extra term $\f{E_0 m}{\alpha_0^2} \delta_{ij}$ appears in the right hand side, but the resulting metric is still non-Euclidean. (From this, it is observed that Newtonian mechanics can be compatible with Euclidean metric only for sufficiently low velocities.)

We observe that the Newtonian Lagrangian which is a consequence of absolute time leads to a metric for space which is inconsistent with Euclidean geometry of space in inertial frames. This observation, conversely, means that Euclidean geometry of space in inertial frames restricts the possible forms of Lagrangian, as will be seen in the following.

\section{Lorentz symmetry as a consequence of invariance of action}\label{S: Lorentz symmetry}
\nin In the previous section, we dealt with properties of Lagrange equations in the extended configuration space. We now restrict our configuration space to the ordinary space so that the extended space becomes identified with spacetime, which could be described by a coordinate system $\{t, q^i\}$. Since we are interested in inertial frames in this section, we may assume that this coordinate system describes a global inertial frame. However, since in all of the derivations that follow only infinitesimal intervals of space and time appear, it suffices that $\{\D t, \D q^i\}$ form a local inertial frame at a certain point $\{t_0, q^i_0\}$ in spacetime. We assume that at this point there is an observer and we call this frame \emph{the observer frame}. We also assume that this observer describes the motion of a free particle relative to his frame and the rest frame of particle is assigned a similar local inertial frame, defined as $\{\D \tau, \D u^i\}$ at the same point in spacetime. We mostly work with this \emph{first} particle, but we occasionally need to assume a \emph{second} particle at rest in the rest frame of the observer. However, when we only mention "particle", we refer to the first particle which is moving relative to the observer.

We also mention that since we consider motions of particles relative to local inertial frames at a fixed point in spacetime, the Lagrangian and Hamiltonian of the particle depend only on the velocity of particle relative to the inertial frame and do not have space or time dependency.

\subsection{Obtaining the general form of the free particle Lagrangian}

\nin We have for the above-mentioned particle

\begin{equation}\label{eq: dS}
\D S = p_i \D q^i - H \D t = - E_0 \D \tau,
\end{equation}

\nin where $E_0$ denotes the rest energy of the particle, and $p_i$ and $H$ denote momentum and energy of the particle relative to the observer. Since the particle has been assumed to have a well-defined rest frame, according to our analysis in section \ref{S: Rest energy - Rest frame}, we must have $E_0 \neq 0$ and this is crucial for the derivations that follow.

Here, $S(t, q^i)$ is a well-defined function whose exact form does not matter. Equation \eqref{eq: dS} expresses infinitesimal differences in this function in terms of space and time intervals in the two frames. Since the particle is at rest in its frame (where momentum is zero), space interval does not appear in the right-hand side.

The equation allows us to obtain equations for \emph{mutual time dilations} between the two frames.\footnote{We use \emph{time dilation} only as a jargon here since at this stage of the analysis we only know that the time intervals are not invariant due to the invariance of action. The same applies to the term \emph{length contraction}.}

To obtain the time dilation of the particle's clock relative to the observer's, we use the equation of motion of the particle, $\D q^i = v^i \D t$, relative to the observer frame, where $v^i$ denotes the velocity of the particle relative to that frame. Using this into equation (\ref{eq: dS}), we obtain $\D t = - \D \tau E_0/L$. To obtain the other time dilation, i.e. the time dilation of the observer's clock relative to the particle's, it suffices to use the equation of motion of the observer's clock in its frame, which is simply $\D q^i = 0$. This yields $\D \tau = \D t H/E_0$.

Here, we have used the action function of the \emph{first} particle to obtain the mutual time dilations between the two frames. Now, we assume the above-mentioned second particle at rest relative to the observer and we apply its action function to obtain the same time dilation relations. If we denote the action function and the rest energy of this particle by $S^\prm$ and $E^\prm_0$, respectively, we can obtain for the same time dilations that $\D t = \D \tau H^\prm/E_0^\prm$ and $\D \tau = - \D t E_0^\prm/L^\prm$, respectively, where $H^\prm$ and $L^\prm$ are Hamiltonian and Lagrangian of the second particle relative to the frame of the first particle.

It follows from the above time dilation relations that

\begin{equation}\label{eq: Mutual time dilations}
\f{H^\prm}{E_0^\prm} = - \f{E_0}{L}, \;\; -\f{E_0^\prm}{L^\prm} = \f{H}{E_0}.
\end{equation}

\nin Here, $H$ must be treated as a function of $v^i$  through the dependence of $p_i$ on $v^i$, that is, we have $H = H(p_i(v^j))$. Similarly, we have $H^\prm = H^\prm(p^\prm_i(v^{\prm j}))$, where $p^\prm_i$ and $v^{\prm i}$ denote, respectively, the momentum and the velocity of the second particle (sitting in the observer's frame) relative to the first particle (moving relative to the observer). An implicit relationship between $v^{\prm i}$ and $v^i$ is implied by equations \eqref{eq: Mutual time dilations}.

Before continuing, we mention that a symmetry of mutual time dilations between the two frames (that is, $\D \tau/{\D t}$ in one time dilation be equal to $\D t/{\D \tau}$ in the other one) requires having $-E_0/L = H/E_0$ and $-E_0^\prm/L^\prm = H^\prm/E_0^\prm$ (which, because of (\ref{eq: Mutual time dilations}), are not independent equations), or simply $HL = - E_0^2$ and $H^\prm L^\prm = - E_0^{\prm 2}$. We show below that this symmetry exists, if we assume Euclidean geometry of \emph{space} in our inertial frames.

Writing (\ref{eq: Mutual time dilations}) in the form of $H^\prm L = - E_0 E_0^\prm = H L^\prm$ and using the relationship between Lagrangian and Hamiltonian, we obtain 

\begin{equation}\label{eq: Fractions series}
\f{p^\prm_i v^{\prm i}}{p_i v^i} = \f{L^\prm}{L} = \f{H^\prm}{H} = - \f{E_0 E_0^\prm}{HL}.
\end{equation}

\nin Now, because of the relationship $L^\prm = -E_0 E_0^\prm/H$, we can write $p^\prm_i v^{\prm i} = v^{\prm i} \di /{\di v^{\prm i}} (-E_0 E_0^\prm/H)$, which using the equality of the first and last fractions in (\ref{eq: Fractions series}) yields

\begin{equation}
v^{\prm i} \f{\di}{\di v^{\prm i}} (\f{1}{H}) = v^i \f{\di L}{\di v^i} \f{1}{HL}.
\end{equation}

We must now use a relationship between the velocities of the two particles, $v^{\prm i}$ and $v^i$, relative to each other. In the standard configuration of coordinate axes, where the corresponding coordinate axes of the two frames are parallel to each other, one finds $v^{\prm i} = - v^i$. From this, we have $v^{\prm i} \di/{\di v^{\prm i}} = v^i \di/{\di v^i}$.\footnote{This applies also to the largest class of configurations of axes where the two relative velocity vectors are related to each other by an arbitrary matrix $R$ in the form $v^{\prm i} = R^i_j v^j$. The importance of this note is that an equation like \eqref{eq: v^i dHL/d v^i = 0} which is supposed to hold in one inertial frame can not depend on the configuration of axes relative to another frame.} Then, we obtain from the recent equation that

\begin{equation}\label{eq: v^i dHL/d v^i = 0}
v^i \f{\di}{\di v^i} (HL) = 0.
\end{equation}

If we employ here the usual assumption of Euclidean geometry of \emph{space} in inertial frames, this equation reduces to $v \di (HL)/{\di v} = 0$, where $v$ is the speed of particle. So, if $v$ is nonzero, we find that $HL$ does not depend on $v$ and therefore has the same value as for $v = 0$, that is $-E_0^2$ ($L(v=0) = - E_0$, $H(v=0) = E_0$). Therefore, in the Euclidean space we have

\begin{equation}\label{eq: HL = - E_0^2}
HL = - E_0^2.
\end{equation}

This is, as mentioned above, \emph{the condition of mutual time dilations symmetry} between the two frames. So, we observe that invariance of action and Euclidean geometry of space provide the symmetry of mutual time dilations between any two inertial frames. 

Using the definitions of Hamiltonian and momentum ($H = p_i v^i - L$, $p_i = \di L/{\di v^i}$), one easily obtains from \eqref{eq: HL = - E_0^2} that

\begin{equation}\label{eq: L^2 - E_0^2}
v^i \f{\di}{\di v^i} (L^2 - E_0^2) = 2 (L^2 - E_0^2).
\end{equation}

\nin In the Euclidean space, this yields $L^2 - E_0^2 = A v^2$, where $A$ is a constant independent of velocity. By redefining this constant as $A = - K E_0^2$, where $K$ is a new constant independent of velocity and the minus sign is for later convenience, the free particle Lagrangian is obtained as

\begin{equation}\label{eq: Lagrangian}
L = - E_0 \sqrt{1 - K v^2}.
\end{equation}

We have chosen the minus sign behind the square root, because the equation must reduce to $L = - E_0$ for $v = 0$. 

From the way $K$ has been introduced, it is not obvious that this constant is independent of $E_0$ or is universal, or even is positive. All we know for the moment is that $K$ must be \emph{nonzero} and \emph{finite} in order for the Lagrangian to make sense. (No implicit dependence of $E_0$ on $K$ is understood from our analysis.) This restriction on $K$ is also implied by the resulting expressions of energy and momentum,

\begin{subequations}\label{eq: p_i, H}
	\begin{align}
	p_i &= E_0 K \f{\delta_{ij} v^j}{\sqrt{1 - K v^2}}, \\
	H &= \f{E_0}{\sqrt{1 - K v^2}} = E_0 \sqrt{1 + \f{p^2}{K E_0^2}} \;\; .
	\end{align}
\end{subequations}

These expressions give more information about $K$. By inserting them into (\ref{eq: dS}), we obtain

\begin{equation}\label{eq: dtau}
\D \tau = \gamma (\D t - K \delta_{ij}v^j \D q^i), \;\; \gamma \equiv \f{1}{\sqrt{1- K v^2}}.
\end{equation}

One can immediately understand from this equation that $K$ does not depend on $E_0$ since none of the other quantities in the equation is assumed to do so. Similarly, $K$ does not depend on $E_0^\prm$ (the rest energy of the second particle) either. These can be shown more concretely, as follows. 

We first note that the time dilation of the particle relative to the observer (obtained from $\D q^i = v^i \D t$) reads as $\D t = \gamma \D \tau$. Had we started from considering the action function of the second particle (at rest relative to the observer), $S^\prm$, we would have obtained the equation $\D t = \gamma^\prm (\D \tau + K^\prm \delta_{ij} v^j \D u^i)$, where $\gamma^\prm = (1-K^\prm v^2)^{-1/2}$ and we have assumed $v^{\prm i} = - v^i$ for the velocity of the first frame relative to the second. Then, for the same time dilation we would have obtained (using $\D u^i = 0$): $\D t = \gamma^\prm \D \tau$. Considering the two expressions of the same time dilation, one obtains $\gamma^\prm = \gamma$ and therefore $K^\prm = K$. Since the rest energies of the two particles, $E_0$ and $E_0^\prm$, can be arbitrarily different, we conclude that not only the constant $K$ is shared between the two frames considered here, but, it is also independent of the particles' rest energies.\footnote{This actually means that the space and time measurements using clocks and meter sticks in inertial frames do not depend on the masses of those clocks and meter sticks. Such a statement may be invalid in the quantum gravity regime.} 

We, however, do not conclude immediately that $K$ is absolutely universal. In fact, we still let $K$ be a property of \emph{the pair} of frames or of the transformation between the pair of frames, meaning that when considering three frames $A$, $B$, and $C$, we let $K$ differ from the transformation between $A$ and $B$ to the transformation between $B$ and $C$.\footnote{We should mention at this point, before completing the derivation of Lorentz transformation, that the invariance of action and Euclidean geometry of space not only provide a symmetry of mutual time dilations between the two frames, as shown previously, but also a \emph{symmetry of mutual longitudinal-length contractions}, which is shown as follows (here, by the term "longitudinal" we mean \emph{parallel to the relative velocity}). Consider the equation $\D S = - E_0 \D \tau = p_i \D q^i - H \D t$ for the particle once more. Also, consider an infinitesimal rod at rest in the frame of particle which is defined by intervals $\D u^i$ and $\D q^i$ in the two frames. Since the observer measures the length of the rod instantaneously ($\D t = 0$), we have: $\D \tau = - \D q^i p_i/{E_0}$. The same procedure can be described using the equation $\D S^\prm = - E_0^\prm \D t = p^\prm_i \D u^i - H^\prm \D \tau$ of the second particle at rest relative to the observer, which using $\D t = 0$ yields: $\D \tau = \D u^i p^\prm_i/H^\prm$. From these two results, and using the expressions of energy and momentum (eqs. (\ref{eq: p_i, H})), we obtain $\gamma \delta_{ij} v^j \D q^i = \delta_{ij} v^j \D u^i$. Similarly by putting a rod in the frame of observer and considering the two equations of action functions we obtain $\gamma \delta_{ij} v^j \D u^i = \delta_{ij} v^j \D q^i$, which considering the previous result shows the symmetry of \emph{longitudinal} length contractions between the two frames. Regarding possible \emph{transversal} length contraction we can not say anything directly using the equation $\D S = p_i \D q^i - H \D t$, since a transversal $\D q^i$ disappears from this equation. What we conclude here from the recent results is that transversal vectors in one frame are transversal in the other frame too.}

\subsection{The general form of the transformation}\label{S: Transformation}
\nin Equation (\ref{eq: dtau}) is one equation of the transformation (in the infinitesimal form) from the observer frame to the rest frame of our first particle. The other equations of the transformation may be written as $\D u^i = A^i_j \D q^i + B^i \D t$, in which, using the equation of motion of the particle relative to the two frames ($\D u^i = 0$ and $\D q^i = v^i \D t$), we have $B^i = - A^i_j v^j$, where $A^i_j = A^i_j(v^k)$. Now, the general form of the transformation equations reads as:

\begin{subequations}\label{eq: Transf.}
\begin{align}
\D \tau  &= \gamma (\D t - K \delta_{ij}v^j \D q^i) \label{eq: Transf. dtau}\\
\D u^i &= A^i_j (\D q^j - v^j \D t),
\end{align}
\end{subequations}

\nin where $\gamma = 1/\sqrt{1 - K v^2}$.

By putting the equation $\D q^i = 0$ and its equivalent $\D u^i = - v^i \D \tau$, in the above equations, one obtains

\begin{equation}\label{eq: A^i_j v^i = gamma v^i}
A^i_j v^j = \gamma v^i.
\end{equation}

In a one-dimensional problem, one immediately finds from this equation that $A = \gamma$. In the case of more than one dimension, one may proceed as follows.

Whatever form the coefficients $A^i_j$ have, in the standard configuration of axes, they must be built functionally from the elements $v^i$, $\delta_{ij}v^j$, and $\delta^i_j$, where $\delta_{ij}$ is the Euclidean metric. One can convince oneself that the most general expression for $A^i_j$ for the standard configuration of axes is $A^i_j(v^k) = \alpha(v^k) \delta^i_j + \beta(v^k) v^i \delta_{jl} v^l$, where $\alpha$ must reduce to $1$ for $v= 0$, if $\beta$ remains finite. Any other suggested form can be reexpressed in this form (adding terms with any number of multiplications of $v^i$, $\delta_{ij}v^j$, and $\delta^i_j$ just leads to redefinition of $\alpha(v^k)$ and $\beta(v^k)$). Then, equation (\ref{eq: A^i_j v^i = gamma v^i}) immediately yields: $\beta = (\gamma - \alpha)/v^2$. 

Therefore, the general form of the coefficients $A^i_j$, without determining $\alpha$, is

\begin{equation}\label{eq: A^i_j}
A^i_j (v^k) = \alpha \delta^i_j + \f{\gamma - \alpha}{v^2} v^i \delta_{jl} v^l, \; \; \alpha = \alpha(v^k).
\end{equation}

For usage in the following subsection, we need to solve the equations (\ref{eq: Transf.}) for $\D t$ and $\D q^i$. From those equations, we obtain $(A^i_j - K \gamma v^i \delta_{jl} v^l) \D q^j = \D u^i + v^i \D \tau$. Multiplying by $\tilde{\alpha} \delta^k_i + \tilde{\beta} v^k \delta_{im} v^m$ and deriving $\tilde{\alpha}$ and $\tilde{\beta}$, we obtain

\begin{subequations}\label{eq: Inv. Transf.}
\begin{align}
\D t  &= \gamma (\D \tau + K \delta_{ij} v^j \D u^i)\\
\D q^i &= \tilde{A}^i_j (\D u^j + v^j \D \tau),
\end{align}
\end{subequations}

\nin where

\begin{equation}\label{eq: tilda(A)^i_j}
\tilde{A}^i_j (v^k) = \f{1}{\alpha} \delta^i_j + \f{\gamma - 1/\alpha}{v^2} v^i \delta_{jl} v^l.
\end{equation}

\subsection{Universality of the constant $K$}\label{S: Universality of K}
\nin We prove the universality of $K$, and at the same time determine $\alpha$ in \eqref{eq: A^i_j}, by taking a third inertial frame into consideration. 

Let us call our previous inertial frames with space-time coordinate systems $\{t, q^i\}$ and $\{\tau, u^i\}$, the frames $A$ and $B$, respectively, and consider a third frame, called $C$, with space-time coordinate system $\{\sigma, w^i\}$. We assume that the velocity of $B$ relative to $A$ is $v^i$ as before, and those of $C$ relative to $B$, and of $C$ relative to $A$ are $v^{\prm i}$ and $v^{\prm\prm i}$, respectively. 

We assume that the transformation between the frames $A$ and $B$ contains the constant $K$ and the function $\alpha$ as before. The transformations from $B$ to $C$ or from $A$ to $C$ will contain an equation in the general form of equation (\ref{eq: Transf. dtau}), though with different values for $K$. (This is a consequence of Lagrangian formalism, not of the principle of relativity (see section \ref{S: Principle of Rel.}) or group property of transformations.) That is, we can write

\begin{subequations}\label{eq: dsigma 1, 2}
	\begin{align}
	\D \sigma  &= \gamma^\prm(v^{\prm k}) (\D \tau - K^\prm \delta_{ij} v^{\prm j} \D u^i), \label{eq: dsigma 1}\\
	\D \sigma  &= \gamma^{\prm\prm}(v^{\prm\prm k}) (\D t - K^{\prm\prm} \delta_{ij} v^{\prm\prm j} \D q^i), \label{eq: dsigma 2}
	\end{align}
\end{subequations}

\nin where $\gamma^\prm(v^{\prm k}) = (1 - K^\prm v^{\prm 2})^{-1/2}$ and $\gamma^{\prm\prm}(v^{\prm\prm k}) = (1 - K^{\prm\prm} v^{\prm\prm 2})^{-1/2}$, and $v^{\prm\prm i}$, the velocity of $C$ relative to $A$, is found using the transformation between $A$ and $B$ in the form of equations \eqref{eq: Inv. Transf.} as

\begin{equation}
v^{\prm\prm i} = \f{\tilde{A}^i_j (v^{\prm j} + v^j)}{\gamma (1 + K \vec{v}.\vec{v}^\prm)} = \f{\f{1}{\alpha} v^{\prm i} + \f{\gamma - 1/\alpha}{v^2} (\vec{v}.\vec{v}^\prm) v^i + \gamma v^i}{\gamma(1 + K \vec{v}.\vec{v}^\prm)}.
\end{equation}

Now, another way of expressing the equation \eqref{eq: dsigma 2} is to use equations \eqref{eq: Transf.} in \eqref{eq: dsigma 1}, which yields

\begin{equation}\label{eq: dsigma indirect}
\begin{aligned}
\D \sigma &= \gamma^\prm \Big\{ \gamma (1 + K^\prm \vec{v}.\vec{v}^\prm) \D t \\
&- [K \gamma v^j + K^\prm \alpha v^{\prm j} + K^\prm \f{\gamma - \alpha}{v^2} \vec{v}.\vec{v}^\prm v^j] \delta_{ij} \D q^i \Big\}.
\end{aligned}
\end{equation}

Comparison between equations \eqref{eq: dsigma 2} and \eqref{eq: dsigma indirect} shows that we must have

\begin{align}
\gamma^{\prm\prm} &= \gamma^\prm \gamma (1 + K^\prm \vec{v}.\vec{v}^\prm)\\
\gamma^{\prm\prm} K^{\prm\prm} v^{\prm\prm j} &= \gamma^\prm [K \gamma v^j + K^\prm \alpha v^{\prm j} + K^\prm \f{\gamma - \alpha}{v^2} \vec{v}.\vec{v}^\prm v^j],
\end{align}

\nin or consequently

\begin{equation}
\begin{aligned}
K^{\prm\prm} \f{1 + K^\prm \vec{v}.\vec{v}^\prm}{1 + K \vec{v}.\vec{v}^\prm} &\Big\{ \f{1}{\alpha} v^{\prm j} + \gamma v^j + \f{\gamma - 1/\alpha}{v^2} (\vec{v}.\vec{v}^\prm) v^j \Big\} =\\
&K^\prm \alpha v^{\prm j} + K \gamma v^j + K^\prm \f{\gamma - \alpha}{v^2} \vec{v}.\vec{v}^\prm v^j.
\end{aligned}
\end{equation}

\nin Here, we have assumed that the $\gamma$ factors are nonzero and finite because the inertial frames are rest frames of massive objects whose velocity, energy, and momentum, as measurable quantities, must make sense.

The recent equation must be an identity for arbitrary velocities $v^j$ and $v^{\prm j}$. As a consequence, the coefficients of the terms $v^{\prm j}$, $v^j$, and $\vec{v}.\vec{v}^\prm v^j$ on both sides must be equal, respectively. From the second terms we obtain

\begin{equation}
K^{\prm\prm} \f{1 + K^\prm \vec{v}.\vec{v}^\prm}{1 + K \vec{v}.\vec{v}^\prm} = K,
\end{equation}

\nin which, since $K$, $K^\prm$ and $K^{\prm\prm}$ do not depend on velocity, means that we must have $K^\prm = K = K^{\prm\prm}$. 

Then, we obtain from the first terms that

\begin{equation}
	\alpha^2 = 1.
\end{equation}

\nin Since the transformation between $A$ and $B$ must reduce to identity in the standard configuration of axes, we must have $\alpha = 1$. 

These results can be extended to any number of frames by induction. So, $K$ not only is not a property of particle as shown before, it can not even be chosen arbitrarily for the transformation between pairs of frames, and therefore is a universal constant to be chosen once for all inertial frames.

\subsection{The sign of $K$}
\nin Knowing that $K$ is a universal constant, we must determine its sign. This final information can be obtained as follows. We first mention a consequence of Lagrangian formalism for velocities of massive particles. If we denote the Lagrangian of a massive free particle relative to two inertial frames by $L(v^i)$ and $L^\prm(v^{\prm i})$, because of the invariance of action, we have  $L^\prm \D t^\prm = L \D t$, or $\sqrt{1 - K v^{\prm 2}} \D t^\prm = \sqrt{1 - K v^2} \D t$, where $t$ and $t^\prm$ are time parameters in the two frames. This equation implies that it is impossible for the velocity of a massive particle to be finite in one frame and infinite in the other. (Infinite velocity is possible only for $K < 0$.) So, $v$ and $v^{\prm}$ are both finite or infinite, depending on what the sign of $K$ allows.

Now, Let us see the same issue from the perspective of the equations of velocity addition:

\begin{equation}\label{eq: Addition of vel.}
\f{\D u^i}{\D \tau} = \f{A^i_j (v^k) (\f{\D q^j}{\D t} - v^j)}{\gamma (v^k) (1 - K \delta_{ij} v^j \f{\D q^i}{\D t})}.
\end{equation} 

\nin Multiplying both sides by $\delta_{il} v^l$, we get the following equation, which is easier to analyze:

\begin{equation}\label{eq: Addition of vel. 2}
\vec{v}.\vec{U}
 = \f{\vec{v}.\vec{V} - v^2}{1 - K \vec{v}.\vec{V}}.
\end{equation} 

\nin Here, $V^i = \f{\D q^i}{\D t}$ and $U^i = \f{\D u^i}{\D \tau}$, and we have assumed that $\gamma(v^k)$ is nonzero and finite for the same reason explained in the previous subsection.

We already know from equations (\ref{eq: Lagrangian}) and (\ref{eq: p_i, H}) that $K$ can not be zero or infinite in order for the Lagrangian formalism to give meaningful results for massive particles. Also, we understand from the same equations and the universality of $K$ that for $K > 0$, there will be an upper limit $c \equiv 1/\sqrt{K}$ for the velocity of massive particles, while for $K < 0$ there will be no such an upper limit on the velocity (There will, however, be an upper limit on the momentum in this case). 

Now, let us assume that $K$ is negative and $V^i$ represents the velocity of a massive particle. Then, since, in this case, there is no restriction on $V^i$ whatsoever, it can be chosen such that we have $\vec{v}.\vec{V} = K^{-1}$, which makes the denominator in the equation \eqref{eq: Addition of vel. 2} vanish. This means that, in the equations of velocity addition, it is possible for the case of negative $K$ that a finite velocity of a massive particle in one frame appears as an infinite velocity relative to another frame, which, regardless of being meaningless physically, is inconsistent with the consequence of invariance of action mentioned above, i.e. the velocity of a massive particle can not be finite in one frame and infinite in another one. On the other hand, such a meaningless result or inconsistency with the invariance of action does not occur for a positive $K$ because in this case, we have $K \vec{v}.\vec{V} \le K v V < 1$ ($V = \sabs{\vec{V}} $). This holds even for the case $V = 1/\sqrt{K}$ although this velocity does not apply to massive particles since, according to (\ref{eq: p_i, H}), it requires a zero rest energy. (For the same reason, we assumed that the relative velocity of the two frames, $v$, can not reach the upper limit $1/\sqrt{K}$ since, as discussed in section \ref{S: Rest energy - Rest frame}, the inertial frames are rest frames of massive objects.)

So, meaningfulness of velocity addition or consistency of velocity addition with invariance of action require $K > 0$. Consequently, the only meaningful possible value for the universal constant $K$ in the Lagrangian formalism lies within the range $0 < K < \infty$.

It may be understood from equations (\ref{eq: p_i, H}) that $1/\sqrt{K}$ is the speed of particles with no rest energy. Although this is correct, but the argument is not correct since all our derivations in this section, including the equations (\ref{eq: Lagrangian}) and (\ref{eq: p_i, H}), were based on the assumption of $E_0 \neq 0$, and, therefore, can not be applied to massless particles. Such particles require an analysis of their own, which is presented in section \ref{S: Massless Particles}.

\section{Massless particles}\label{S: Massless Particles}
\nin As discussed in section \ref{S: Rest energy - Rest frame}, invariance of action implies that particles with no rest energy can not have well-defined inertial rest frame. From this, it is understood that the velocity of such a particle can not be reached by any moving observer, otherwise the rest frame of observer would be a rest frame for the particle too. Moreover, from equations \eqref{eq: p_i, H}, it follows that for $E_0 = 0$ we can only have $v = \f{1}{\sqrt{K}} = c$, which is the upper limit for the speed of massive objects and reference frames. However, since the entire analysis in the previous section was based on the assumption $E_0 \neq 0$, the case of massless particles requires an analysis of its own.

A proper description of motions of such particles is achieved by considering motions for which the Lagrangian $L$ vanishes. This means that for such a motion we must have the stationarity condition $\delta \int (p_i v^i - H(t, q^i, p_i)) \D t = 0$, alongside the additional condition $p_i v^i - H(t, q^i, p_i) = 0$. From the first, we obtain the ordinary Hamilton equations for the motion, and then using the second one, we obtain

\begin{equation}
	H = p_i \f{\di H}{\di p_i}.
\end{equation}

In the Euclidean space, this equation reduces to 

\begin{equation}
	H = p \f{\di H}{\di p},
\end{equation}

\nin where $p$ is the magnitude of momentum. The last equation has only one solution, which is

\begin{equation}\label{eq: H = C p}
	H = C p,
\end{equation}

\nin where $C$ is a constant independent of the momentum and energy. Using Hamilton equations, one easily obtains for this motion that

\begin{equation}
	v = C,
\end{equation}

\nin where $v$ is the speed of particle. Therefore, the speed of a particle in Euclidean space in a motion for which the Lagrangian vanishes is a constant, independent of energy and momentum. In such a motion, $p = 0$ corresponds to $H = 0$, and therefore, the particle does not have a rest energy. It, therefore, according to section \ref{S: Rest energy - Rest frame}, can not have a well-defined inertial rest frame.

Now, this analysis is applicable in any inertial frame because, due to the equation \eqref{eq: El} and $\El$ being a scalar, the equation $L = 0$ holds in all inertial frames. This, however, does not directly show that $C$ is the same in all inertial frames, and the equality of $C$ with the speed constant $c$ in the Lorentz transformation is not understood immediately from the above analysis. For these, we require comparison between different inertial frames and therefore employing the Lorentz transformation between them. At the moment, we just know that in order for the energy to be finite, $C$ must be finite in all inertial frames. 

We may divide the possible values of $C$ into three classes, $C < c$, $C = c$, and $C > c$, where $c$ is the universal speed constant in the Lorentz transformation. The case $C > c$ is discarded because according to the law of velocity addition, it allows a finite velocity $C$ relative to one inertial frame to appear as an infinite velocity relative to another inertial frame, which is physically unacceptable. Moreover, an infinite velocity makes $H$ infinite in equation \eqref{eq: H = C p}. The case $C < c$ is also discarded because we already know from equations \eqref{eq: Lagrangian} and \eqref{eq: p_i, H} that any $v < c$ can be the speed of a massive particle and this means that no $v < c$ can be attributed to a particle with no rest energy, otherwise, according to the law of velocity addition, the rest frame of the massive particle will also be the rest frame of the massless particle which can not have a rest frame. Therefore, the only possibility for the energy-independent velocity $C$ of massless particles in Euclidean space is the universal constant $c$ appearing in the Lorentz transformation between inertial frames. 

\section{Some remarks regarding the derivation of Lorentz transformation}\label{S: Remarks}

\subsection{Genuine role of dynamical considerations}
\nin An important point regarding the derivation of Lorentz symmetry from Lagrangian formalism is to consider how genuinely the derivation depends on relationships that exist due to dynamical considerations, and not due to purely \emph{geometric} ones in the usual sense. 

Our basic equation in that derivation was

\begin{align}
\D S = - E_0 \D \tau = p_i \D q^i - H \D t,
\end{align}

\nin which was applied to the motion of a free particle. If one defines $F = H/E_0$ and $G_i = - p_i/E_0$, this equation can be re-expressed as the spacetime geometric relationship

\begin{align}
\D \tau = F \D t + G_i \D q^i, \label{eq: dtau F, G_i} 
\end{align}

\nin where the coefficients are functions of relative velocity. 

At first sight, it may look like that, in the previous sections, we have just been dealing with geometric relationships in the disguise of dynamical quantities. However, one can easily observe that dynamical considerations provide relationships between the coefficients in this equation which are not understood (at least very easily) from purely geometrical relationships. While geometric considerations (including the assumption of the universality of the speed of light in the standard special relativity) provide \emph{algebraic} relationships between these coefficients, dynamical considerations provide \emph{differential} relationships between them.

From the definitions of momentum and Hamiltonian, one obtains $p_i = \f{\di L}{\di v^i} = \f{\di}{\di v^i}(p_j v^j - H)$ or $v^j \f{\di p_j}{\di v^i} = \f{\di H}{\di v^i}$, which in terms of $F$ and $G_i$ defined above is 

\begin{eqnarray}\label{eq: F & G_i diff. eq.}
\f{\di F}{\di v^i} = - v^j \f{\di G_j}{\di v^i} .
\end{eqnarray}

From purely geometric considerations, including the assumption of universality of the speed of light, it is not obvious why such a special \emph{differential} relationship must exist between $F$ and $G_i$ (of course this is without deriving the Lorentz transformation first!). By taking derivative from \eqref{eq: dtau F, G_i} with respect to $v^j$ and keeping $\D t$ and $\D q^i$ fixed, one finds that \eqref{eq: F & G_i diff. eq.} means that $\D \tau$ will be minimum when $v^i$ is such that one has the relationship $\D q^i = v^i \D t$ between space and time intervals, that is, for the motion of particle or reference frame. This is not a simple assumption to think of, before deriving the transformation (or even after that!) while \eqref{eq: F & G_i diff. eq.} is a simple equation with a simple interpretation in terms of dynamical quantities. It is just a consequence of the definitions of momentum and Hamiltonian and a trivial relationship which directly does not have to say something specially meaningful about the nature of space and time.

On the other hand, this trivial differential relationship between momentum and Hamiltonian leads straightforwardly to the determination of $F$ and $G_i$. It suffices to add to that equation the condition of mutual time dilations symmetry between the two frames, $HL = - E_0^2$, which in terms of $F$ and $G_i$ reads as $F + G_i v^i = 1/F$ (in our approach, this symmetry is a consequence of invariance of action in the Euclidean space; in special relativity, it is an expectation based on the assumptions of universal light-speed and equivalence of inertial frames). Then, one obtains using \eqref{eq: F & G_i diff. eq.} that

\begin{equation}
G_i = - \f{1}{F^2} \f{\di F}{\di v^i}, \;\;\; v^i \f{\di F^2}{\di v^i} = 2 F^2 (F^2 -1),
\end{equation}

\nin with solutions $F^2 = (1 - K v^2)^{-1}$ and $G_i = - K \delta_{ij} v^j (1 - K v^2)^{-1/2}$. 

This analysis clearly demonstrates the crucial role of differential relationships such as \eqref{eq: F & G_i diff. eq.}, as consequences of dynamical considerations, in determining the coefficients in the equations of transformation.

\subsection{Explicit absence of gravity in inertial frames}
\nin In our approach, it is easy to understand that one should not seek a theory of gravity within the framework of special-relativistic kinematics. 

Unlike in special relativity, where inertial frames are treated as abstract Cartesian coordinate systems moving relative to each other and Lorentz symmetry is derived for such abstract systems, in our approach, we derived this symmetry from the consideration of action functions of particles sitting in inertial frames, and, as discussed in section \ref{S: Rest energy - Rest frame}, such particles are necessarily massive. Therefore, in our approach, inertial frames are never abstract coordinate systems, but are rest frames of massive objects. Nevertheless, in obtaining Lorentz transformation, we neglected any gravitational effect of these objects on each other. In the language of general relativity, we neglected any tiny deviations of spacetime from flatness (within local or global inertial frames) due to the presence of energy and momentum of particles.

This is also evident from the fact that, in section \ref{S: Lorentz symmetry}, we assumed two distinct action functions, $S$ and $S^\prm$, in $4$ dimensions for the particles sitting in the two considered inertial frames, instead of assuming one common action function for two particles in their ($7$-dimensional) extended configuration space.

The observation is that Lorentz symmetry is a consequence of Lagrangian formalism by neglecting gravitational effects of moving bodies and therefore is expected to be applicable only in situations where gravitational effects of particles are negligible. In other words, in our approach, one explicitly expects the inapplicability of Lorentz symmetry in the presence of gravitational effects of particles.

This not only rules out the possibility of formulating gravity within the framework of special-relativistic kinematics, but in terms of which, it becomes reasonable that the standard techniques of quantum field theory do not work for gravity. The dispersion relation $E = \sqrt{E_0^2 + p^2 c^2}$ applied in quantum field theory is a consequence of neglecting gravitational effects of particles and can not be used in the situations that those effects become effective, for example for particles with energies comparable with Planck energy.

\subsection{Local inertial frames - Time dilation relative to non-inertial frames - Gravitational time dilation}\label{S: Local meaning}
\nin Although our discussion was concentrated on the derivation of Lorentz symmetry for inertial frames, which covers special relativity, but the formalism goes beyond that and covers also the foundations of general relativity.

Since the derivation of Lorentz symmetry in section \ref{S: Lorentz symmetry} was automatically for infinitesimal spacetime intervals, the considered inertial frames do not need to be global frames. The approach, therefore, applies to local as well as global inertial frames.

For the time dilation of the particle relative to the observer in that section we obtained, by considering the action function of the particle, that $\D t = - \D \tau \, E_0/L$. In that section we assumed that both the observer and the particle are in inertial frames, which ended in the special-relativistic equation $\D t = \D \tau (1 - K v^2)^{-1/2}$. 

However, the equation $\D t = - \D \tau \, E_0/L$, by itself, only requires that the particle be in an inertial frame, not necessarily the observer, and a local inertial frame would suffice for this matter since the equation is in infinitesimal form. This means that the same equation can be applied to some other situations as well. One such situation is when the particle is not freely moving according to the observer (that is, it is under interaction), but, the rest frame of the particle remains at least locally inertial. A second situation that the same time dilation relation applies is when the particle is in a global inertial frame but the observer is accelerating. In other words, when the particle is in a local or global inertial frame but the observer is not, the observer still observes a time dilation for the particle, determined by the Lagrangian of particle relative to the observer reference frame, no matter what the state of motion of the observer is.

The second situation mentioned here describes the time dilation of a free particle relative to an accelerating observer, while the first one, according to the experiences underlying the Einstein equivalence principle, happens for particles moving in a gravitational field. Therefore, the time dilation equation mentioned above describes also \emph{gravitational} time dilation. In this situation, the Lagrangian $L$ describing the motion of the freely falling particle is a general function $L(t, q^i, v^i)$ expressible as $ L = - E_0 \sqrt{ g_{00} + 2 g_{0i} v^i + g_{ij} v^i v^j } $ because of \eqref{eq: El} and \eqref{eq: Metric}. The metric $g_{\mu \nu}$ is in general Finsler, but, it reduces to a Riemann metric because of Lorentz symmetry in (local) inertial frames which provides a pseudo-Euclidean metric in those frames.\footnote{One point to be noted in this regard is that the inertia-gravity equivalence principle is implicit in the assumption of inertial frames. In other words, when we assume that some frames are inertial we are implicitly employing the equivalence principle. If there was no such a property for gravity, it would have been impossible to have even a single (local) inertial frame in the universe, since different objects in one reference frame would have different accelerations relative to another object and therefore assigning a common rest frame to different objects would have been impossible.}

\subsection{Insufficiency of an apparently short argument}
\nin Based on the appearance of the equation $\D S = p_i \D q^i - H \D t$, one may suggest a much shorter argument and derivation, compared with what was presented in section \ref{S: Lorentz symmetry}, for concluding Lorentz symmetry as a consequence of invariance of action. We first mention the argument and then argue that it is not correct.

The equation $\D S = p_i \D q^i - H \D t = p_\mu \D q^\mu$, where $p_\mu = (-H, p_i)$ and $q^\mu = (t, q^i)$, can be recast in the form $\D S = \eta_{\mu \nu} p^\nu \D q^\mu$, where $\eta_{\mu \nu} = \textup{diag}(-1, 1, 1, 1)$ and $p^\nu = (H, p_i)$. The right-hand side of the equation $\D S = p_\mu \D q^\mu$ is the same in any arbitrary coordinate system of any type. However, if one chooses a coordinate system in which $p_i$ and therefore $p^\mu$ appear as vectors like $\D q^i$ and $\D q^\mu$, that is a Cartesian coordinate system, then a transformation $\D q^\mu = \Lambda^\mu_\sigma \D q^{\prm \sigma}$ will be accompanied by a transformation $p^\mu = \Lambda^\mu_\sigma p^{\prm \sigma}$ and the matrix $\eta_{\mu \nu}$ will be the same in all such coordinate systems, i.e. one has $\eta_{\mu \nu} \Lambda^\mu_\sigma \Lambda^\nu_\lambda = \eta_{\sigma \lambda}$. It is already well-known that a transformation of Cartesian coordinates which preserves this $\eta_{\mu \nu}$ is a Lorentz transformation. It is not hard to understand that a velocity constant $c$ must be introduced in the derivation to compensate for the difference between units of space and time coordinates and it will finally appear as an invariant velocity under the transformation.

The point to be noted about this argument is that the minus sign behind $H$ in the equation of $\D S$ is due to the \emph{definition} of Hamiltonian. If we defined $H = L - p_i v^i$, it would disappear and we must have obtained a transformation preserving a matrix $\eta_{\mu \nu} = \textup{diag}(1, 1, 1, 1)$! So, appealing directly to the appearance of equation $\D S = p_i \D q^i - H \D t$ to conclude Lorentz symmetry as a consequence of invariance of action does not provide a sound argument. For that matter, one has to employ a more careful argument like the one presented in section \ref{S: Lorentz symmetry}.

It is, however, interesting to examine what happens if we change the definition of Hamiltonian to $H = L - p_i v^i$.

The basic equation would be $\D S = p_i \D q^i + H \D t = E_0 \D \tau$, and the symmetry of mutual time dilations would yield $HL = E_0^2$ leading to the same equation \eqref{eq: L^2 - E_0^2} for $L$, but this time with the solution $L = E_0 \sqrt{1 - K v^2}$, since now $L$ must reduce to $E_0$ for $v^i = 0$. From this Lagrangian, one obtains

\begin{subequations}
	\begin{align}
	p_i &= - E_0 K \f{\delta_{ij} v^j}{\sqrt{1 - K v^2}}, \\
	H &= \f{E_0}{\sqrt{1 - K v^2}},
	\end{align}
\end{subequations}

\nin leading to the same equation \eqref{eq: dtau}. All the arguments for the derivation of Lorentz symmetry would be as before and would lead to the same Lorentz transformation with $K > 0$.

So, the above-mentioned short argument is not correct, since changing the definition of Hamiltonian by a minus sign, which changes the appearance of the equation of $\D S$, does not affect the Lorentz transformation derived. This is, of course, pleasing because we do not expect fundamental physics depend on our choices or definitions. 

We, however, observe that for the modified definition of $H$, momentum of the particle is on the opposite direction relative to its velocity, and if we assume $E_0 < 0$ to compensate that, this makes $H < 0$. The latter is meaningless if we keep the usual notion of kinetic energy ($H - E_0$) as the work done on the particle ($\int v^i \D p_i$) to set it in motion. For both definitions of $H$, the condition $H-E_0 > 0$ requires $E_0 K > 0$, and since $K$ is universal and positive, we require $E_0 > 0$ for that matter.

In short, if one defines $H = \lambda_H (p.v - L)$, $K = \lambda_K \sabs{K}$, and $L = \lambda_L \sabs{E_0} \sqrt{1 - K v^2}$, where $\lambda_H^2 = \lambda_K^2 = \lambda_L^2 = 1$, then the conditions $E_0 \ge 0$ (required for $H - E_0 > 0$) and $p.v \ge 0$ (required for $\int v^i \D p_i > 0$) as physical requirements lead respectively to $\lambda_H \lambda_L = -1$ and $\lambda_K \lambda_L = -1$, and therefore, $\lambda_H = \lambda_K$. Then, since consistency of velocity addition with Lagrangian formalism requires $\lambda_K = +1$, as mentioned in section \ref{S: Lorentz symmetry}, we must have $\lambda_H = +1$, i.e. we must define $H$ as usual in order to keep the mentioned conditions. This justifies the usual definition of Hamiltonian. 

In other words, changing the definition of Hamiltonian does not affect Lorentz symmetry, but, Lorentz symmetry alongside physical requirements ($E_0 \ge 0$, $p.v \ge 0$) determines the suitable definition of Hamiltonian.

\subsection{On the principle of relativity (equivalence of inertial frames )}\label{S: Principle of Rel.}
\nin In our derivation of Lorentz transformation, universality of the speed of massless particles was derived as a final consequence of invariance of action (in Euclidean space). Therefore, the second postulate of special relativity has not been used. However, it appears that even the principle of relativity, the first postulate of special relativity, has not been employed in the derivations as a principle distinct from or more fundamental than the formalism of classical dynamics. 

One may confirm by inspection that in no part of the derivations in sections \ref{S: Lorentz symmetry} and \ref{S: Massless Particles} we appealed to equivalence of inertial frames to conclude something that is not derivable directly or by a sequence of derivations from Lagrangian formalism. All parts of the derivations, including spatio-temporal symmetry of inertial frames in Euclidean space, were systematic consequences of invariance of action in the Lagrangian formalism, 

Not appealing to the equivalence of inertial frames as a distinct assumption besides Lagrangian formalism is evident from the fact that we did not assume the universality of either the constant $K$ (in the transformation equations in section \ref{S: Lorentz symmetry}) or the constant $C$ (the speed of massless particles in section \ref{S: Massless Particles}) based on the assumption of equivalence of inertial frames, but we got those as systematic consequences of Lagrangian formalism. Likewise, we did not appeal to the equivalence of inertial frames for the derivation of $\alpha$ in the equations of the transformation, which could be done by making a comparison between equations \eqref{eq: Transf.} and their inverse \eqref{eq: Inv. Transf.}.

Two points should be added in this regard. First, the very applicability of Lagrangian formalism in different inertial frames is not an application of the principle of relativity but a consequence of the Lagrangian formalism itself. This is simply because, following our discussion in section \ref{S: Invariance of action} and the equivalence between equations \eqref{eq: Lagrange eqs of L} and \eqref{eq: Lagrange eqs of El}, Lagrangian formalism is intrinsically invariant under arbitrary change of coordinate system in the extended configuration space (\emph{spacetime} in this case). It, therefore, applies in different inertial frames too. Second, the assumption of Euclidean geometry of space in different inertial frames, employed for the derivation of Lorentz transformation, should not be considered as an application of the principle of relativity. As the analysis of section \ref{S: Reduction} suggests, this assumption is equivalent to the statement that the frame is inertial, and is not a special or distinct law of nature to be held in different inertial frames. In fact, without this assumption the frame would not be the subject of the principle of relativity at all. 

So, in our approach, the derivations were all consequences of the Lagrangian formalism and they did not require the assumption of the equivalence of inertial frames as an additional assumption or an assumption more fundamental than the formalism of classical dynamics.

Now, we may even argue that the equivalence of inertial frames (\emph{with respect to Lorentz symmetry}) can be considered as the consequence of the formalism. As far as one is concerned with classical particle dynamics, this equivalence is implied by Lagrangian formalism. However, since the derived Lorentz symmetry is a spatio-temporal symmetry, this equivalence automatically applies also to any field theory that is expressible in terms of space and time coordinates, using differential equations. This fairly includes all known applications of Lorentz symmetry in classical and quantum field theory. Therefore, for the known applications of Lorentz symmetry, the equivalence of inertial frames can be considered as a systematic consequence of Lagrangian formalism.

\section{Reduction of Finsler metric to Riemann and the meaning of the Euclidean geometry of space in inertial frames}\label{S: Reduction}
\nin The assumption of Euclidean geometry of space in inertial frames was a fundamental assumption for the derivation of Lorentz transformation in section \ref{S: Lorentz symmetry}. Here, we discuss the relationship between the Euclidean geometry of space and inertiality of the frame.

In section \ref{S: Lorentz symmetry}, we showed two things without making reference to the geometry of space. First, using the equation $\D S = p_i \D q^i - H \D t$ and the notion that the particle is at rest in its rest frame, we showed that the condition for symmetry of mutual time dilations between two inertial frames is equivalent to the equation $HL = - E_0^2$. Second, by considering action functions of two such particles, we obtained (equation (\ref{eq: v^i dHL/d v^i = 0})):

\begin{equation}\label{eq: v^i dHL/d v^i = 0 2}
v^i \f{\di}{\di v^i} (HL) = 0.
\end{equation}

This equation has two solutions: either $HL$ is a constant, or it is a homogeneous function of degree zero in $v^i$.

In section \ref{S: Lorentz symmetry}, we showed that the assumption of Euclidean geometry of space yields $HL = - E_0^2$ as the only solution for that equation (for arbitrary $v^i$). Since all the rest of derivations up to the Lorentz transformation and universality of the speed of massless particles followed from the equation $HL = - E_0^2$, we can say that \emph{Euclidean geometry of space in inertial frames is the most fundamental empirical fact underlying Lorentz symmetry}; it is certainly more fundamental than the universality of the speed of massless particles, which, as was shown, does not need to be postulated as an empirical fact or a fundamental assumption about the nature of space and time. 

In the following, we consider consequences of equation (\ref{eq: v^i dHL/d v^i = 0 2}), if we do not impose the Euclidean geometry of space as an assumption.

The first thing we note is that $HL$ as a homogeneous function of degree zero in $v^i$ is inconsistent with one assumption that has been used in the derivation of equation  (\ref{eq: v^i dHL/d v^i = 0 2}), i.e. the assumption that the particle is at rest in its inertia frame. This is because any function $f(v^i)$ which is homogeneous of degree zero in $v^i$ can be written as $f(\f{v1}{v^3}, \f{v^2}{v^3}, 1)$, in $3$-space. Such a function does not depend on the magnitude of the velocity components, but depends on the direction of velocity. As a consequence, it is ill-defined or in fact undefined for the vanishing velocity (all components of $v^i$ vanish). This is inconsistent with the assumption that the particle is at rest in its inertial frame, because in this case we have $H(v^i = 0) = E_0$ and $L(v^i = 0) = - E_0$, and therefore we have the well-defined value $HL = - E_0^2$ for $v^i = 0$. So, we observe that the mentioned assumption does not allow $HL$ to be a function of $v^i$ at all, and therefore, according to which, $HL$ must be $-E_0^2$ for arbitrary $v^i$. This means that the equation $HL = - E_0^2$ as the acceptable solution of equation (\ref{eq: v^i dHL/d v^i = 0 2}), which was obtained as a consequence of imposing the assumption of Euclidean geometry of space in section \ref{S: Lorentz symmetry}, actually does not require this assumption about the geometry of space and is a consequence of the assumption that the particle is at rest in its inertial rest frame.

Below, we argue that the fundamental reason behind this observation is that the statements \emph{space is Euclidean} and \emph{the frame is inertial} are equivalent.

To observe this, we consider the consequences of equation (\ref{eq: v^i dHL/d v^i = 0 2}) for the geometry of spacetime, if we do not impose the assumption of Euclidean geometry of space.

The Euclidean geometry of space in inertial frames, as was shown, provides Lorentz symmetry between inertial frames, implying a Minkowski spacetime in those frames, and therefore, a general Riemann metric in arbitrary coordinate systems in spacetime. However, as mentioned in section \ref{S: metric}, in the Lagrangian formalism, the metric of spacetime can be of Finsler type, in general, instead of strictly Riemann type.

Since equation (\ref{eq: v^i dHL/d v^i = 0 2}) was obtained before making any reference to geometry, it is supposed to hold even when the geometry of \emph{spacetime} is Finsler. 

From the definition of Finsler metric in (\ref{eq: Metric}) and of $\El$ in (\ref{eq: El}) we have 

\begin{equation}\label{eq: L in terms of g}
L = \alpha_0 \sqrt{g_{00} + 2 g_{0i} v^i + g_{ij} v^i v^j},
\end{equation}

\nin where, $v^i = \f{\D q^i}{\D t}$, and $g_{\mu \nu} (q^\lambda, \dot{q}^\lambda)$ are homogeneous functions of degree zero in $\dot{q}^\mu = \f{\D q^\mu}{\D \tau}$, and $\tau$ is an arbitrary parameter. Comparison with \eqref{eq: Lagrangian} shows that $\alpha_0$ can be identified with $- E_0$. Because of the zero-degree homogeneity of the metric, we can write $g_{\mu \nu} (q^\lambda, \f{\D q^\lambda}{\D \tau}) (\f{\D \tau}{\D t})^0 = g_{\mu \nu}(q^\lambda, \f{\D q^\lambda}{\D \tau} \f{\D \tau}{\D t}) = g_{\mu \nu}(q^\lambda, 1, v^i)$ and therefore $g_{\mu \nu}$ can be considered as functions of $v^i$. If they do not depend on $v^i$, they will represent a Riemann spacetime. There are no specific conditions on the functionality of $g_{\mu \nu}$ in terms of $v^i$ at the moment.

From the Lagrangian (\ref{eq: L in terms of g}), one obtains

\begin{equation}\label{eq: HL in Finsler}
HL = - \alpha_0^2 \; f_{(0)}(v^i),
\end{equation}

\nin where,

\begin{equation}\label{eq: f_(0)}
\begin{aligned}
f_{(0)}(v^i) = g_{00} + g_{0i} v^i &- \f{1}{2} v^i \f{\di g_{00}}{\di v^i} - v^i \f{\di g_{0j}}{\di v^i} v^j \\
&- \f{1}{2} v^i \f{\di g_{jk}}{\di v^i} v^j v^k.
\end{aligned}
\end{equation}

\nin Because of (\ref{eq: v^i dHL/d v^i = 0 2}), $f_{(0)}(v^i)$ must be homogeneous of degree zero in $v^i$, hence the subscript '$(0)$'. 

We have derived (\ref{eq: HL in Finsler}) from the most general Lagrangian (\ref{eq: L in terms of g}) in Finsler spacetime, without actually enforcing the condition (\ref{eq: v^i dHL/d v^i = 0 2}). However, similar to what was done in section \ref{S: Lorentz symmetry}, we may do the reverse and obtain the Lagrangian from (\ref{eq: HL in Finsler}), knowing that $f_{(0)}(v^i)$ must be homogeneous of degree zero in $v^i$. In other words, we may derive the consequences of equation (\ref{eq: v^i dHL/d v^i = 0 2}) for the form of the Lagrangian (\ref{eq: L in terms of g}), and obtain the general form of Lagrangian for a free particle in Finsler spacetime.

We will have

\begin{equation}
v^i \f{\di}{\di v^i} (L^2 - \alpha_0^2 f_{(0)}(v^k) ) = 2 \; (L^2 - \alpha_0^2 f_{(0)}(v^k)),
\end{equation}

\nin which shows that $L^2 - \alpha_0^2 f_{(0)}(v^k)$ is homogeneous of degree $2$ in $v^i$. So, the general form of the free particle Lagrangian in Finsler spacetime reads as

\begin{equation}\label{eq: L f_(0), f_(2)}
L = \alpha_0 \sqrt{f_{(0)}(v^i) + f_{(2)}(v^i)},
\end{equation}

\nin where $f_{(2)}(v^i)$ is an undetermined function homogeneous of degree $2$ in $v^i$. $f_{(0)}(v^i)$ and $f_{(2)}(v^i)$ must be expressed in terms of the metric components.

Using the equation $\El = L \f{\D t}{\D \tau}$ (equation \eqref{eq: El}), we obtain

\begin{equation}\label{eq: El f_(0), f_(2)}
\begin{aligned}
\El &= \alpha_0 \sqrt{f_{(0)}(v^i) (\f{\D t}{\D \tau})^2 + f_{(2)}(v^i) (\f{\D t}{\D \tau})^2} \\
&= \alpha_0 \sqrt{f_{(0)}(\dot{q}^i) \dot{t}^2 + f_{(2)}(\dot{q}^i)},
\end{aligned}
\end{equation}

\nin where $\dot{q}^\mu = \f{\D q^\mu}{\D \tau}$ and we have used $f_{(2)}(v^i) \dot{t}^2 = f_{(2)}(\dot{q}^i)$ because of the second-degree homogeneity of $f_{(2)}$ in its arguments. Using this Lagrangian in the definition of Finsler metric (equation \eqref{eq: Metric}) to obtain $g_{\mu \nu}$, we obtain 

\begin{equation}
\begin{aligned}
g_{00} &= f_{(0)}(\dot{q}^i) \\
g_{0i} \dot{q}^i &= \Big(\f{\di f_{(0)}}{\di \dot{q}^i} \dot{t}\Big) \dot{q}^i = 0 \\
g_{ij} \dot{q}^i \dot{q}^j &= \Big(\f{1}{2} \f{\di}{\di \dot{q}^i}\f{\di f_{(0)}}{\di \dot{q}^j} \dot{t}^2 + \f{1}{2} \f{\di}{\di \dot{q}^i}\f{\di f_{(2)}}{\di \dot{q}^j} \Big) \dot{q}^i \dot{q}^j \\
&= f_{(2)}(\dot{q}^i)	
\end{aligned}
\end{equation}

\nin and therefore

\begin{equation}\label{eq: f_0 g rels.}
\begin{aligned}
f_{(0)}(v^i) &= g_{00}, \\
g_{0i} v^i &= 0 \\
f_{(2)}(v^i) &= g_{ij} v^i v^j.
\end{aligned}
\end{equation}

These results could also be read off directly from \eqref{eq: El f_(0), f_(2)}. These equations show that $g_{00}$ and $g_{ij}$, which are homogeneous of degree zero in $\dot{q}^\lambda$ but supposedly arbitrary functions of $v^i$ according to the definition of Finsler metric, are homogeneous of degree zero in $v^i$, as a consequence of equation (\ref{eq: v^i dHL/d v^i = 0 2}). 

Because of (\ref{eq: f_0 g rels.}), equation \eqref{eq: HL in Finsler} reduces to 

\begin{equation}\label{eq: HL = - alpha_0^2 g_{00}}
HL = - \alpha_0^2 \, g_{00}.
\end{equation}

As discussed above, the statement that $HL$ is homogeneous of degree zero in $v^i$, instead of being the constant $-E_0^2$, is inconsistent with the notion that the particle is at rest in its inertial frame. Now, equation (\ref{eq: HL = - alpha_0^2 g_{00}}) implies that $g_{00}$ being homogeneous of degree zero in $v^i$, too, is inconsistent with the notion of having a particle at rest in its inertial frame.

What we observe here is \emph{a conflict between Finsler geometry of spacetime and the assumption that the particle is at rest in its inertial rest frame}. 

The conflict is resolved only for $g_{00}$ not being a function of $v^i$ at all. Since in transformation between different frames, the components of metric get mixed to each other, it follows that all components of metric must be independent of $v^i$ in order for $g_{00}$ to become independent of $v^i$ in all such frames. In other words, the conflict is resolved only if the metric is reduced to a Riemann metric.

So, we observe that Finsler spacetime is incompatible with this simple notion that \emph{a particle is sitting at rest in its inertial frame}, and since the concept of inertial frame means that the particle can be freely at rest, this implies that \emph{Finsler spacetime is incompatible with the notion of inertial frame}. (We remember that the assumption that the particle is freely at rest in a frame and will remain so --- if the frame were hypothetically global --- means that the frame is inertial.) In other words, the concept of inertial frame has meaning and possibility only in a Riemann spacetime.\footnote{In a Riemann spacetime, then, for $v^i = 0$, we obtain $L = \alpha_0 g_{00}^{1/2}$ and we must have $L = - E_0$. $\alpha_0$ is identifiable with $-E_0$ and $g_{00}$ can be made $1$ be redefining the time coordinate in the local frame. Considering Euclidean geometry of space in inertial frames, the general form of free particle Lagrangian in inertial frame will be obtained as $L = - E_0 \sqrt{1 - K\delta_{ij} v^i v^j}$, and for determining the properties of $K$, one must follow an analysis similar to the one given in section \ref{S: Lorentz symmetry}.}

Now, since in a Riemann spacetime, one can always find local Cartesian frames with Euclidean metric for space, it follows that the concept of \emph{inertial frame} is not separable from the assumption of Euclidean geometry of \emph{space} in such frames and in fact is \emph{equivalent} to that assumption. In other words, when we assume that a reference frame is inertial, we are assuming that geometry of space in that frame is Euclidean.

This point is specially important for any derivation of Lorentz symmetry, including ours in section \ref{S: Lorentz symmetry}, where one assumes that geometry of space in inertial frames is Euclidean. From the above analysis it becomes clear that the assumption of Euclidean space is not an assumption added to the assumption of inertial frame. It has the same meaning.

Another aspect of the above issue with Finsler metric is as fallows. In Finsler spacetime, along with losing the concept of inertial frame, we also lose the essential \emph{symmetry} property of freely moving frames, even if we somehow give the meaning of inertiality to such frames in Finsler spacetime. Since the condition of symmetry of mutual time dilations between two frames, in which, particles can be freely at rest, is equivalent to the equation $HL = - E_0^2$ and this equation, as discussed above, does not hold in Finsler spacetime, we conclude that \emph{in a general Finsler spacetime, there is no symmetry of mutual time dilations between freely moving reference frames}. Even if we somehow generalize the concept of \emph{inertial} frame for Finsler spacetime, we still encounter the problem that the frames do not have spatio-temporal symmetry relative to each other. From a phenomenological point of view, calling freely-moving frames in Finsler spacetime equivalent is problematic, even though the general formalism of classical dynamics still applies in them.

\section{Energy-independent velocity as a general type of motion in classical dynamics}\label{S: Energy-independent vel.}
\nin In section \ref{S: Massless Particles}, we concluded the universality of the speed of massless particles as a (final) consequence of the invariance of action and Euclidean geometry of space, and the derivation of Lorentz symmetry in section \ref{S: Lorentz symmetry} did not require the assumption of universal light-speed as a fundamental kinematical principle about the nature of space and time. One observation which is consistent with this conclusion and supports our proposition of maintaining a unifying view regarding relativity and classical dynamics is that the concept of energy-independent velocity, which is considered as the characteristic of massless particles in the standard formulation of relativity, is actually a more general concept in classical dynamics and in fact a type of motion appearing even in massive systems, and the universal speed of massless particles is just one example of such a motion. Masslessness or universality of speed are not fundamental requirements in order for a system to have an energy-independent motion in terms of appropriate canonical coordinates.

In the following subsections, we first discuss the general definition and properties of motions with an energy-independent velocity, and then, by mentioning some examples, we show how this general type of motion may find specific properties in different physical conditions.

\subsection{General definition and properties of motions with an energy-independent velocity}
\nin Let us assume that we are describing a motion for which the Lagrangian vanishes. As mentioned also in section \ref{S: Massless Particles}, this means that we must have the stationarity condition $\delta \int (p_i v^i - H(t, q^i, p_i)) \D t = 0$ alongside with the additional condition $p_i v^i - H(t, q^i, p_i) = 0$ for the extremal path. The first equation yields the ordinary Hamilton equations, and, then, from the second equation one obtains

\begin{equation}\label{eq: H = p_i di H/ di p_i}
p_i \f{\di}{\di p_i} H = H,
\end{equation}

\nin which means that $H$ must be homogeneous of degree one in $p_i$. This, in turn, yields

\begin{equation}
p_i \f{\di}{\di p_i} v^j = 0,
\end{equation}

\nin which means that $v^j$ is homogeneous of degree zero in $p_i$ in this motion.

These properties imply that multiplication of momentums $p_i$ by any numerical factor multiplies $H$ by the same factor, and vice versa, and $v^j$ will remain unaffected.

So, in general, if a motion is such that the Lagrangian vanishes, the velocities $v^j$ of such a motion are independent of the energy and momentum of the particle. Since motion of light and massless particles, discussed in section \ref{S: Massless Particles}, is an example of this type of motion, we may call such a motion a \emph{light-like motion}.

Since properties of such a motion depend on the vanishing of Lagrangian, and Lagrangian in general is not canonically-invariant, this concept is not canonically-invariant. In other words, a motion which is light-like (energy-independent) in one canonical system can be an ordinary (energy-dependent) motion in another canonical system, and vice versa.

In section \ref{S: Massless Particles}, we discussed such a motion for one-particle systems in the ordinary Euclidean space, and we do not repeat it here. We explore some other examples of such a motion in various physical conditions.

\subsection{In Riemann spacetime}\label{S: Light-like in Riemann}
\nin In a Riemann spacetime, the equation $L = 0$, or equivalently $\El = 0$ due to \eqref{eq: El}, leads to $g^{\mu \nu} p_\mu p_\nu = 0$. From this equation, one easily obtains

\begin{equation}\label{eq: Light-like H in Riemann}
H = \f{g^{0i}p_i}{g^{00}} \pm \sqrt{\bigg(\f{g^{0i}p_i}{g^{00}}\bigg)^2 - \bigg(\f{g^{ij}p_i p_j}{g^{00}}\bigg)},
\end{equation}

\nin which is first-degree homogeneous in $p_i$ and satisfies (\ref{eq: H = p_i di H/ di p_i}). By assuming $g_{0i} = 0$, it is clear that the motion described by this Hamiltonian is possible only in a pseudo-Riemannian spacetime (Lorentzian signature).

As far as the transformation between observers is assumed to be a point transformation of spacetime, this light-like motion is \emph{observer-independent}, since equation $\El = 0$ is coordinate-independent in spacetime.

Such a motion may have specific properties in various examples of Riemannian spacetime.

In Schwarzschild geometry, one has

\begin{equation}
H = \sqrt{ \Big( 1 - \f{2GM}{r} \Big) \Big[ \Big( 1 - \f{2GM}{r} \Big) p_r^2 + \f{p_\theta^2}{r^2} + \f{p_\varphi^2}{r^2 \sin^2 \theta} \Big]}.
\end{equation}

Radial light-like motion ($p_\theta = 0$, $p_\varphi = 0$) is possible on both sides of the horizon, while circular light-like motion ($p_r = 0$) is possible only outside the horizon. Moreover, using Hamilton equations one easily finds that the circular motion is possible only for $ r = \f{3GM}{c^2}$, as is already well-known.\\

In a geometry, like Kerr, where $g_{0i} \neq 0$, there are two sets of light-like motions possible corresponding to the $\pm$ signs in \eqref{eq: Light-like H in Riemann}.\\

In a flat FLRW geometry, one finds

\begin{equation}
H = \f{c}{a} (\delta^{ij} p_i p_j)^{1/2} = \f{c}{a} p,
\end{equation}

\nin where $a(t)$ is the scale factor. Since, according to Hamilton equations, $p_i$, canonical momentums corresponding to co-moving coordinates, are constants of motion, this equation means that speed of light-like motion in the co-moving frame decreases by expansion as $c/a(t)$. This equation is, however, usually interpreted as the decrease of physical momentum by expansion as $p/a(t)$, although, the first interpretation seems more relevant from the viewpoint of Hamilton equations and the chosen coordinate system.

\subsection{In anisotropic space}
\nin Equation \eqref{eq: H = p_i di H/ di p_i} by itself can have varieties of solutions in more than one dimension. That equation is mathematically too general and finds different types of solution in different physical conditions. In Euclidean space which is an isotropic space, it has only one solution, which is $H = C p$, where $C$ is a constant independent of energy and momentum, and $p$ is the magnitude of momentum. However, in an anisotropic environment, the equation

\begin{equation}
H = C^i p_i,
\end{equation}

\nin $C^i$ being distinct constants independent of energy and momentum, is a solution to equation \eqref{eq: H = p_i di H/ di p_i}. In this situation, velocity of the particle performing light-like motion is different in different directions.

\subsection{Harmonic and anti-harmonic motions as light-like motions}
\nin The fact that the angular velocity of classical oscillator does not depend on the energy of the system can be related to the issue of light-like motions and be interpreted as a consequence of performing a light-like motion in angular canonical system. It is known \cite{c: Gold} that the Hamiltonian

\begin{equation}
H = \f{p^2}{2m} + \f{1}{2} k q^2  
\end{equation}

\nin can be written as

\begin{equation}\label{eq: H = omega theta}
H = \omega P_\Theta;  \;\; \omega = \sqrt{\f{k}{m}},
\end{equation}

\nin by the canonical transformation

\begin{equation}\label{eq: p,q to P_Theat and Theta}
p = \sqrt{2 P_\Theta m\omega} \cos \Theta, \;\; q = \sqrt{\f{2 P_\Theta}{m\omega}} \sin \Theta.
\end{equation}

In the canonical system of $\Theta$ and $P_\Theta$, the Lagrangian of the oscillator vanishes and the angular velocity ($\dot{\Theta} = \omega$) becomes independent of the angular momentum $P_\Theta$ and of the energy $H$, as expected. Therefore, an \emph{ordinary} motion like harmonic oscillation of a \emph{massive} system in the \emph{ordinary} space appears as a \emph{light-like} motion in an \emph{appropriate canonical system}.

A similar analysis can be done for anti-harmonic motion ($H = p^2/{2m} - kq^2/2$), replacing triangular functions in (\ref{eq: p,q to P_Theat and Theta}) by hyperbolic functions.

\subsection{Motion of \emph{light} as harmonic/anti-harmonic motion}
\nin Linear energy-independent motion of \emph{light} in the ordinary space can be expressed as an energy-dependent motion, such as harmonic oscillation, in terms of appropriate canonical coordinates. We assume that we are dealing with a one dimensional motion described by canonical linear variables $q$ and $p$. In order to express the motion of light as a harmonic oscillation, we have to introduce a length parameter $\lambda$ such that $q/\lambdabar$ ($\lambdabar = \lambda/{2\pi}$) can be considered as the phase of the oscillator (that is, $\Theta$ of the previous subsection be identified with $q/\lambdabar$). In other words, we are reinterpreting the linear coordinate $q$ as an angular coordinate in order to use a transformation like (\ref{eq: p,q to P_Theat and Theta}). We now have $\omega = \dot{\Theta} = c/\lambdabar$ and from the equation $H = \omega P_\Theta = pc$, we obtain $P_\Theta = \lambdabar p$. Now, using the canonical transformation

\begin{equation}
P = \sqrt{2pc\alpha} \cos \f{q}{\lambdabar}, \;\; Q = \lambdabar \sqrt{\f{2p}{c\alpha}} \sin \f{q}{\lambdabar},
\end{equation}

\nin where $\alpha$ is a free parameter, the Hamiltonian $H = pc$ is expressed as

\begin{equation}
H = \f{P^2}{2\alpha} + \f{1}{2} \f{c^2 \alpha}{\lambdabar^2} \, Q^2.
\end{equation}

Here, in terms of coordinate $Q$, the motion of light is momentum-dependent and energy-dependent as for an ordinary oscillator (in a Cartesian coordinate): 

\begin{equation}
\dot{Q} = \f{P}{\alpha} = \pm \sqrt{\f{2H}{\alpha} - \f{c^2}{\alpha \lambdabar^2} Q^2}.
\end{equation}

Similarly, one can represent the motion of light as an anti-harmonic motion.

\subsection{Generality for systems with one degree of freedom}
\nin For systems with one degree of freedom, any ordinary (energy-dependent) motion can be expressed as an energy-independent motion in a canonical coordinate system. If $H(q, p)$ is the Hamiltonian of such a system in terms of a canonical coordinate $q$ for which the motion is energy-dependent, one can define $P(q, p) = \f{1}{\omega} H(q, p)$ as the canonical momentum corresponding to a canonical coordinate $Q=Q(q, p)$, whose exact functionality in terms of $q$ and $p$ does not matter. From the new Hamiltonian $H^\prm(Q, P) = H(q, p) = \omega P$, one obtains $\dot{Q} = \omega$ and $L^\prm = P\dot{Q} - H^\prm = 0$.

Of course, such a mathematical generality is not always accompanied by physical significance. Here, one may simply set $\omega = 1$ and define $H(q, p)$ as the new canonical momentum and mathematically obtain a trivial energy-independent constant velocity with no physical significance. It is understood that it is only for few systems, such as harmonic oscillator and massless particles, that the energy-independent velocity has direct physical meaning and usefulness. 

\subsection{Light-like motion as a general concept}
\nin As the above examples show, the notion of energy-independent velocity, for the generalized meaning of the term \emph{velocity}, is a general type of motion in classical dynamics. In terms of the coordinates of the ordinary space and for one-particle systems this motion is possible only for massless particles, but, in some canonical coordinates, it appears even for massive particles. 

Despite the above statement in the previous subsection regarding the physical significance of light-like motions in its general mathematical possibility, it seems that, in principle, there is no reason to assume that such a motion with physical meaning and application can not be found in various systems.

In principle, any system of particles (and even fields) can have its own light-like motions in a proper configuration space. Such motions, however, may not appear energy-independent when expressed in terms of the variables of the ordinary space. But, the fact that the system does have an energy-independent generalized velocity in a proper configuration space might still be identifiable, similar to the situation of harmonic oscillator, whose energy-independent angular velocity is identifiable even in terms of linear coordinates and momenta. 

In general, such a property means that the motion (or evolution) of system in terms of certain canonical coordinates does not depend on the energy of the system. Such a system can, therefore, move or evolve with lowest possible energy.

The physical importance of this feature, if such motions could actually be realized in various physical systems, requires no emphasis. Observation of such a behavior in multi-particle or more complicated systems would provide empirical evidence for the necessity of maintaining a unified view regarding relativity and classical dynamics, and would have serious implications for our understanding of relativity, spacetime, and dynamical systems.

However, one point to be noted here is that this entire analysis is based on classical dynamics. It will be interesting if a quantum mechanical counterpart for this classical concept can be developed.

\section{Local meaning of the universality of the speed of massless particles and the possibility of variable speed of light}
\nin The derivation of Lorentz symmetry in section \ref{S: Lorentz symmetry} was essentially based on working with infinitesimal intervals of time and space. As discussed in the beginning of that section, the considered local inertial frames must exist at a certain point $\{t_0, q^i_0\}$ in spacetime. As a consequence, the \emph{universality} of the constant $K$ discussed in that section (corresponding to $1/c^2$ in the conventional notation) can only mean independence of $K$ of all such local inertial frames existing at that point in spacetime. Because of the local nature of the analysis, one can not infer that $K$ is a constant for the entire (curved) spacetime. The most straightforward assumption we are allowed to make about $K$, purely based on our approach, is that $K$ is a \emph{local constant}, specific to the point $\{t_0, q_0\}$ in spacetime. Nothing in our approach necessitates or implies that $K$ is constant for the entire spacetime. 

This implies that, due to the local nature of the analysis, the speed of massless particles relative to local inertial frames is allowed to be a property of the point $\{t_0, q^i_0\}$ and differ from one point to another point in the spacetime, in terms of a global coordinate system. This means that the speed of massless particles in local inertial frames is allowed to be a scalar field on the spacetime, instead of being a strictly constant quantity for the entire spacetime. 

Absolute constancy of the speed of light measured by inertial observers is an assumption in general relativity and there is no mechanism for its variation with space and time in the standard formalism of the theory. But, some researchers argue that variable speed of light might help in solving some problems in theoretical physics \cite{c: VSL01,c: VSL02,c: VSL03} and it might have evidences in observations \cite{c: VSL01}. 

Global constancy of $c$, if provable at all by the Lagrangian formalism, is not implied by our presented analysis. The fact that Lorentz symmetry is derivable from invariance of action, without requiring the assumption of universal light-speed as a fundamental kinematical property of space and time, can be considered as a theoretical foundation for variable speed of light cosmology.

What theoretically supports the idea of variable speed of light is that the motions of massless particles in spacetime are most fundamentally characterized by energy-independent velocity, not by the stronger condition of an absolutely constant speed (relative to local inertial frames) in the entire spacetime. As observed from section \ref{S: Light-like in Riemann}, the most general characterization of light-like motion (of massless particles) in a Riemann spacetime is by the vanishing of Lagrangian. It is only for the special situation of considering the motion relative to (local) inertial frames that this property is translated to a universal speed, and this speed, as discussed above, is only a local constant (universality here means sameness for all massless particles relative to local inertial frame). The condition of vanishing Lagrangian, by itself, does not require the \emph{additional} assumption that that speed, as measured by local inertial observers, is globally a constant quantity for the entire spacetime. In other words, global constancy of the local speed of massless particles is not implied by the vanishing of Lagrangian. 

If that speed changes with space and time, the motion still remains energy-independent for the entire trajectory of the particle. This means that arrival time experiments can only prove the energy-independent velocity of light signals with different energies and not the absolute constancy of their speed relative to local inertial frames throughout the trajectory of the light signal. So, such experiments can not disprove the possibility of variable light-speed with space and time.

\section{Why only one kinematical scale?}\label{S: One Scale}
\nin Unlike the approach of special relativity, we did not begin with postulating a fundamental kinematical velocity scale, i.e. the speed of light. But, nevertheless, we ended up with such a scale as a consequence of invariance of action and Euclidean geometry of space. All the special-relativistic effects that are consequences of the universality of the speed of light in special relativity are attributed to the invariance of action in our approach, including the very existence of that velocity scale, and that velocity is the only kinematical scale we encounter in our approach. However, in recent years, it has been suggested in the quantum gravity research, such as in the so-called doubly special relativity approach, that the short-scale/high-energy structure of spacetime may be governed by \emph{two} kinematical scales, not just one. It has been suggested that the Planck length or Planck energy might have a kinematical role similar to that of the speed of light \cite{c: QG01,c: DSR01,c: DSR02,c: DSR03,c: DSR04}. It is necessary to have some comments on these ideas, based on our discussions and derivations in the previous sections. 

The above mentioned investigations have shown that realization of such a two-scale kinematics is possible, at least in momentum space, using non-linear representations of Lorentz symmetry \cite{c: DSR01,c: DSR02,c: DSR03,c: DSR04}. This realization requires non-linearity (or non-flatness) of momentum space \cite{c: DSR05,c: DSR06} and it has been argued that this non-linearity is accompanied by a non-commutativity of spacetime coordinates \cite{c: DSR07}. Based on the correspondence between quantum mechanical commutators and Poisson brackets, this requires non-vanishing Poisson brackets between spacetime coordinates.

Although we did not employ a Hamiltonian formalism in the extended configuration space\footnote{The Hamiltonian corresponding to $\El$ vanishes identically and can not provide a suitable Hamiltonian formalism.} and made use only of the properties of Lagrange equations in the configuration and extended configuration spaces, it seems reasonable to assume that spacetime coordinates would remain commutative if we had to develop a proper Hamiltonian formalism for the extended space, which, according to the above mentioned investigations, implies lack of a second scale.

The absence of a second scale corresponding to energy and momentum in our approach is very well justified and is a consequence of one of our assumptions. As mentioned in section \ref{S: Rest energy - Rest frame}, a direct consequence of Hamilton-Jacobi equations and invariance of action is the equation $p^\prm_\nu = p_\mu {\di q^\mu}/{\di q^{\prm \nu}}$, indicating that $p_\mu$ are $1$-forms spanning a linear space. This linearity justifies right away the absence of an energy-momentum scale. But the more fundamental reason behind this absence seems to be our assumption of \emph{point transformations} preserving Lagrange equations, which is also apparent from the transformation equation of $p_\mu$. In other words, assuming the transformations between observers as point transformations of spacetime makes the momentum space a linear space.

We note that the assumption of point transformations was fundamental for our derivation of Lorentz symmetry in section \ref{S: Lorentz symmetry} since we assumed two different local inertial frames at a given point $\{ t_0, q_0\}$ in spacetime and consistency of the derivation requires that this assumption be independent of the global coordinate system describing the situation (existence of the two inertial frames at a single point in spacetime).

The connection between point transformations of spacetime and linearity of momentum space and the consequent lack of an energy-momentum scale can be understood from a different angle. It has been argued that curvature of momentum space leads to a relativity of locality \cite{c: DSR08,c: DSR09}. The latter can be interpreted as a manifestation of \emph{non-point transformations} in spacetime, in which, the transformation of spacetime coordinates depends not only on the spacetime coordinates themselves but also on the first or higher-order derivatives of those coordinates. For example, one may have $q^{\prm \mu} = q^{\prm \mu} (q^\nu, \f{\D q^\nu}{\D \tau})$, where $\tau$ is a scalar parameter. Such transformations map different vectors at the same point in spacetime to different vectors at different points in spacetime. As a consequence, two particles with different $4$-velocities which are at one point in spacetime according to the description of one observer would appear at different spacetime points according to the description of another observer.

The conclusion from our derivations is that as long as the spacetime transformations are assumed to be point transformations, the momentum space will be linear with no kinematical scale, and such a scale requires non-point transformations as a necessary condition.

One other point to be noted regarding the above-mentioned two-scale kinematics is related to the suggested implementations of such a kinematics using Finsler geometry \cite{c: DSR10,c: DSR11,c: DSR12}. As discussed in section \ref{S: Reduction}, one problem with Finsler spacetime is that freely-moving frames (at one point in spacetime) do not have spatio-temporal symmetry relative to each other, and this poses a problem for the equivalence between such frames. So, if the whole point of the doubly special relativity approach is the proposition of a new kinematics and a new transformation for inertial frames which are supposed to be equivalent \cite{c: DSR01}, realization of this program using Finsler geometry is problematic.

\section{On the relationship between kinematics and dynamics}
\nin In special relativity, as in any other parts of physics, the kinematics is assumed first and based on which proper forms of dynamics are developed, which includes constructing a proper form of classical mechanics (relativistic mechanics). In this paper, we did the reverse and obtained the special-relativistic kinematics in inertial frames as a consequence of classical dynamics. The meaning of this work is not to revoke the kinematical character from Lorentz symmetry as it clearly serves as a kinematical symmetry, specially for field theories developed in inertial frames. However, the notion of the derivation of a so-called kinematical symmetry from the general framework of classical dynamics affects the perceived relationship between kinematics and dynamics.

Even as implied by the word "kinematics", we usually take kinematical assumptions as given, without expecting a foundation in anything dynamical, and in fact we use such assumptions to build or develop theories of dynamics. So, kinematical assumptions serve as foundations for theories of dynamics, and it seems that such theories are impossible to develop without making kinematical assumptions whatsoever. Theories of dynamics are built upon kinematical assumptions. The very structure of the formalism of classical dynamics is based on many kinematical assumptions; assumptions which are not consequences of classical dynamics, as has been developed so far, such as the concepts of particle, space, time, particle trajectories and lack of fundamental uncertainty, and so on.

On the other hand, the very possibility of deducing a symmetry, which serves as a kinematical assumption for a very broad range of phenomena, from a theoretical framework of dynamics provides the notion that the relationship between kinematics and dynamics can be a two-way relationship. Not only kinematics serves as foundation for theories of dynamics, it, by itself, can be a consequence of some theories of dynamics. In other words, \emph{dynamical phenomena can produce kinematical relationships}; relationships between theoretical elements of the dynamical theories which remain unaffected for a broad range of dynamical phenomena. Here, the $1 + 3$ Riemann structure of spacetime which is a kinematical assumption in general relativity, valid for a very broad range of phenomena, finds a foundation in classical dynamics.

This point of view, that kinematical assumptions can be consequences of dynamical considerations, seems to be the utmost extension of the general-relativistic concept of assigning dynamical nature to space and time. This dynamical nature not only includes spacetime being affected by the energy content of matter but also contains the very nature and structure of spacetime, and it may have to be extended to any symmetry which at the moment is assumed as a purely-kinematical symmetry, such as gauge symmetries of quantum field theory.

By this explanations, it seems that our approach removes a \emph{conceptual} shortcoming in the standard formulation of general relativity, in which, despite the essential dynamical nature of space and time, Lorentz symmetry in inertial frames is assumed as a purely-kinematical property of spacetime with no dynamical origin or character whatsoever. Here, this symmetry appears as a consequence of classical dynamics and finds a dynamical foundation. 

The dynamical origin of Lorentz symmetry makes the theoretical possibility of variable speed of light (massless particles) less surprising. On the other hand, if empirical evidences show the absolute constancy of the speed of light, not just its local universality, then our presented approach based on classical dynamics would be incomplete unless we provide a proper theoretical explanation of that observation based on classical dynamics or any other theory of dynamics.

\section{Concluding remarks}
\nin A new point of view regarding the foundations of relativity and relativistic kinematics has been introduced, according to which, the current Einsteinian theory of relativity, which is a classical (non-quantum) theory, should be unified with the formalism of classical dynamics. This point of view has theoretical advantages, compared with the standard point of view which defines Lorentz symmetry as an intrinsic property of space and time, with no dynamical foundation. The approach seems as a conceptual improvement in that it provides a dynamical origin for Lorentz symmetry, which adds another aspect to the essential dynamical nature of space and time in relativity.

It should be observed that our approach provides simplicity in two aspects. First, relativistic kinematics and relativistic mechanics are obtained together and there is no need to build a relativistic mechanics after defining relativistic kinematics as in the usual way. In this approach, Lagrangian is derived rather than being properly chosen. Second, neither of the two fundamental assumptions of special relativity, i.e. the principle of relativity and the universality of the speed of light, are required in our approach as assumptions, but appear as consequences. The approach elucidates the fundamental role of the Euclidean geometry of space in inertial frames as the ultimate empirical foundation of Lorentz symmetry, and demonstrates the equivalence between Euclidean geometry of space and inertiality of the frame.

Another point is that our derivation of Lorentz transformation shows a redundancy of assumptions in the conventional "relativistic Lagrangian mechanics." In the latter, one makes two assumptions, the assumption of Lorentz symmetry and the assumption that action must be a Lorentz-scalar. The derivation of Lorentz symmetry as a consequence of invariance of action demonstrates the redundancy of the first as an assumption. 

Although, unlike the approach adopted in special relativity, we did not begin with assuming a kinematical scale of velocity, we ended up with such a scale as a consequence of the formalism of classical dynamics. No any other scale appeared in our approach, specially no scale of energy or momentum. This result is a consequence of our assumption of point transformations in spacetime which makes the momentum space a linear space. Adopting non-point transformations, according to which, the transformation of spacetime can depend on first or higher order derivatives, would be a necessary condition for obtaining such an energy-momentum scale, and this conclusion seems compatible with the result of some investigations that show that curvature of momentum space implies a relativity of locality.

Our derivation of Lorentz Symmetry is fundamentally local, in the sense that it is based on an assumption of locality: spacetime point transformations. It is also inherently local in the sense that it yields the symmetry locally such that the speed of massless particles (relative to local inertial observers) can vary with space and time globally. This may be considered as a theoretical foundation for variable speed of light cosmology.

A consequence of our suggested point of view regarding relativity and relativistic kinematics, which seems important to us, is the observation that energy-independent velocity is a general concept in classical dynamics and is observed even in massive objects in appropriate canonical coordinates. Possible observations of such a behavior in various and complicated systems would have consequences for our understanding of spacetime and dynamical systems and would be an empirical evidence for our point of view regarding unification of relativity with classical dynamics.

\end{document}